\documentclass[aps,pra,preprint,groupedaddress]{revtex4-1}
\usepackage{graphics,amsmath,graphicx,color,natbib}

\begin{document}

\title{Echoes and Revival Echoes in Systems of Anharmonically Confined Atoms}
\author{M. Herrera}
\email[]{mherrer1@umd.edu}
\author{T. M.  Antonsen}
\author{E. Ott}
\affiliation{Dept. of Physics, and IREAP, University of Maryland, College Park, Maryland 20742, USA}
\author{S. Fishman}
\affiliation{Physics Dept., Technion-Israel Institute of Technology, Haifa 32000, Israel}
\date{\today}
\begin{abstract}
We study echoes and what we call `revival echoes' for a collection of atoms that are described by a single quantum wavefunction and are confined in a weakly anharmonic trap.  The echoes and revival echoes are induced by applying two, successive temporally localized potential perturbations to the confining potential, one at time $t=0$, and a smaller one at time $t=\tau$.  Pulse-like responses in the expectation value of position $\langle x (t)\rangle$  are predicted at $t \approx n\tau$ ($n=2,3,\dots$) and are particularly evident at $t \approx 2\tau$.  While such echoes are familiar from previous work, a result of our study is the finding of `revival echoes'. Revivals (but not echoes) occur even if the second perturbation is absent.  In particular, in the absence of the second perturbation, the response to the first perturbation dies away, but then reassembles, producing a response at revival times $mT_x$ ($m=1,2,\dots$).  The existence of such revivals is due to the discreteness of the quantum levels in a weakly anharmonic potential, and has been well-studied previously.  If we now include the second perturbation at $t=\tau$, we find temporally localized responses, \emph{revival echoes}, both \emph{before} and after $t\approx mT_x$, e.g., at $t \approx m T_x-n \tau $ (pre-revival echoes) and at $t \approx mT_x+n\tau$, (post-revival echoes), where $m$ and $n$ are $1,2,\dots$ .  One notable point is that, depending on the form of the perturbations, the `principal' revival echoes at $t \approx T_x \pm \tau$ can be much larger than the echo at $t \approx 2\tau$.  We develop a perturbative model for these phenomena, and compare its predictions to the numerical solutions of the time-dependent Schr\"odinger Equation.  The scaling of the size of the various echoes and revival echoes as a function of the symmetry of the perturbations applied at $t=0$ and $t=\tau$, and of the  size of the external perturbations is investigated. The quantum recurrence and revival echoes are also present in higher moments of position, $\langle x^p (t)\rangle $, $p>1$.   Recurrences are present at $t \approx mT_x/j$, and  dominant  pre and post-revival echoes occur at fractional shifts of $\tau$, i.e. $t\approx(mT_x \pm \tau)/j$,  where the $m=1,2,\dots$ and the integer values of $j$ are determined by $p$.  Additionally, we use the Gross-Pitaevskii Equation to study the effect of atom-atom interactions on these phenomena.  We find that echoes and revival echoes become more difficult to discern as the size of the second perturbation is increased and/or as the atom-atom interactions become stronger. 
\date{today}
\begin{description}
\item[PACS numbers]
03.65.-w, 03.65.Ge
\end{description}
\end{abstract}

\maketitle

\section{Introduction}
In both classical and quantum systems with nonlinearity, a distribution of oscillation frequencies leads to dephasing of the response to a temporally localized system perturbation and to damping in physical observables of the response.  After this response damps away, application of another external perturbation can induce subsequent build up of phase coherence, causing later temporally localized responses, called \emph{echoes}. Examples have been investigated in nuclear magnetic resonance (spin echoes) \cite{Hahn1950}, coupled oscillators \cite{Ott2008}, plasma physics \cite{Gould1967,Malmberg1968,Ott1970}, cavity quantum electrodynamics \cite{Morigi2002,Meunier2005}, and cold atom systems \cite{Bulatov1998,Piovella2003,Buchkremer2000, Maneshi2008,Andersen2003,Elyutin2005}.  Additionally, in quantum mechanical systems, another phenomena, the so called \emph{revivals} due to the discreetness of the energy eigenstates, can occur \cite{Raithel1998,Pitaevskii1997,Robinett2004}.  Revivals are also present in systems with a Jaynes-Cummings type interaction, and have been studied both theoretically \cite{Eberly1980,Buzek1995}, and experimentally in trapped-ion \cite{Meekhof1996}, cavity QED \cite{Brune1996}, and circuit QED \cite{Hofheinz2008} systems.  In particular, revivals are macroscopic responses that can result due to reconstruction of phase coherence in the absence of the second perturbation.    Previous work has investigated, both experimentally and theoretically, revival \cite{Raithel1998}, and echo phenomena in trapped atomic systems induced by changes in depth \cite{Bulatov1998}, or translations of \cite{Buchkremer2000, Maneshi2008}, the confining potential.

Here we consider echo phenomena induced by two impulsive, successive external perturbations applied to a collection of atoms in a weakly anharmonic trap.  For example, the type of external perturbation that we consider in most detail is the application of two translations to the trapping potential, one at time $t=0$, and a smaller translation at time $t=\tau$.  We study the response of the expectation value of position $\langle x (t)\rangle$.  In the case of a Bose-Einstein condensate confined by the pondermotive force of a laser beam, translations of the potential can be realized by switching of the laser beam position, or by a change in phase of two interfering beams \cite{Buchkremer2000}.  As in previous work \cite{Buchkremer2000,Bulatov1998, Maneshi2008},  this method results in the creation of echoes, e.g., at $t\approx 2\tau$, and in the quantum mechanical revival described in \cite{Raithel1998,Pitaevskii1997,Robinett2004} at, e.g., $t\approx T_x$.  In addition, however, we also find quantum `revival echoes' occurring both \emph{before} and after the revival, e.g.,  at $t \approx T_x \pm \tau$ and $t \approx T_x\pm 2\tau$.  Furthermore, we find that these revival echoes can have much larger amplitudes than, say the echo at $t \approx 2 \tau$ (conditions for this to be the case are discussed in Sec. \ref{sym}).  In Sec. \ref{u0dll1} we develop a quantum mechanical, perturbative model (in the limit of small anharmonicity and small second displacement) for echoes and revival echoes.  We then compare these results to the numerically solved, time dependent Schr\"odinger Equation.  We study the behavior of echoes and revival echoes at different choices of anharmonicity and time delay $\tau$, as well as how the size of the various echoes scale with the size of the external perturbations.  In Sec. \ref{sym} we discuss what effect the spatial symmetry of the external stimuli has on the echo phenomena, in particular the echoes' relative sizes.  In Sec. \ref{xp} the quantum recurrence and revival echoes are investigated in higher moments of position, $\langle x^p (t)\rangle $, $p>1$.   Recurrences are present at $t\approx mT_x/j$, and  dominant  pre and post-revival echoes occur at fractional shifts of $\tau$, i.e. $t\approx(mT_x \pm \tau)/j$,  where $m=1,2,\dots$ and the integer values of $j$ are determined by $p$.  Lastly, we model atom-atom interactions in our system using the Gross-Pitaevskii Equation, and study the effect of interactions on the revival echo phenomena (Sec. \ref{u01_d1}).

The previous references most closely related to our work are the Refs. \cite{Bulatov1998,Piovella2003,Buchkremer2000, Maneshi2008,Andersen2003,Elyutin2005} on echoes in systems of confined atoms, none of which find our revival echoes.  Reference \cite{Bulatov1998} presents a theoretical treatment of echoes in confined, cold atoms induced by successive changes in the depth of the confining potential.  Reference \cite{Piovella2003} theoretically and numerically studies the echo phenomena in a Bose-Einstein condensate where the dephasing is reversed with an external optical potential.  Reference \cite{Buchkremer2000} reports on experimental observations of echoes induced by sudden shifts of the potential well realized by changing the relative phasing of interfering laser beams, while Ref. \cite{Maneshi2008} experimentally and numerically compares the effectiveness of translations with different temporal profiles in creating echoes.  Reference \cite{Andersen2003} experimentally observes an echo in a collection of trapped atoms induced by applying a short microwave $\pi$ pulse, where the echo is measured via Ramsey spectroscopy.  Reference \cite{Elyutin2005} numerically implements a Bose-Hubbard model to describe a collection of atoms in an optical lattice and their echo response to pulses of radiation in the presence of atom-atom interactions.

\section{Model and Numerical Results}
\label{u0dll1}

We begin by looking at echo phenomena present in the absence of atom-atom interactions.  We study the quantum evolution of an initially Gaussian state in a weakly anharmonic, one-dimensional trap,

\begin{align}
i \frac{\partial}{\partial t } \psi & = (H_0 + H_1) \psi, \\
H_0 &= -\frac{1}{2} \frac{\partial^2}{\partial x^2} + \frac{1}{2} x^2, \\
H_1 &=\frac{1}{4} \beta x^4,
\end{align}
where $\beta$ quantifies the anharmonicity of the trap, with $\beta \ll 1$. We use harmonic units, $x= \bar{x}/(\sqrt{\hbar/m\omega_0}) $, $t=\omega_0\bar{t}$, where $\bar{x}$ and $\bar{t}$ are the unnormalized units, $\omega_0$ is the frequency of the harmonic oscillator, and $m$ is the mass.  The procedure will be the following.  
\begin{enumerate}
\item Begin with a Gaussian wavepacket centered $x=0$.
\item At $t=0$ translate the state by an amount $d_1$, $x\rightarrow x+d_1$.
\item At $t=\tau$ translate the state again in $x$ by an amount $d_2$.
\end{enumerate}

Later (Sec. \ref{sym}) we will discuss another type of perturbation that is not a translation, and we will find that the spatial symmetry of the perturbations can have important effects.  In order to most clearly stimulate echoes, revivals, and revival echoes, we will initially take  $d_2 \ll 1$.  We first seek the frequencies $\omega_n$ of our anharmonic Hamiltonian, $H=H_0+H_1$, for $H_1$ (i.e., $\beta$) small.  Expanding $\omega_n$ to second order in $\beta$, we write


\begin{align}
\label{freqall}
\omega_n=\omega_n^{(0)}+\delta \omega_{n}^{(1)} +\delta \omega_{n}^{(2)} .
\end{align}
%
The corrections to the frequencies are conveniently calculated using the creation/ annihilation formalism of the harmonic oscillator.  

\begin{align}
\hat{x}&=\sqrt{\frac{1}{2}}(\hat{a}^\dagger + \hat{a}) ,\label{xdef}\\
\hat{p}&=i\sqrt{\frac{1}{2}}(\hat{a}^\dagger - \hat{a}), \\
\hat{a} |n \rangle &= \sqrt{n} |n-1 \rangle ,\\
\hat{a}^{\dagger} |n \rangle &= \sqrt{n+1} |n+1 \rangle ,
\end{align}

Taking $H_1=(\beta/16) (\hat{a}^\dagger + \hat{a})^4 $, perturbation theory \cite{Sakurai1994} yields 
\begin{align}
\label{freq1}
\delta \omega_n ^{(1)}= \langle n |H_1 |n \rangle = \frac{\beta}{16} (3+6n(1+n)),
\end{align}

\begin{align}
\delta \omega_n ^{(2)} & =\sum_{k \neq n} \frac{ | \langle n |H_1| k \rangle |^2}{(n+1/2)-(k+1/2)}.  \notag \\
&= \frac{\beta^2}{256} \bigg\{ \left[2n(1-2n)^2(n-1)\right] \Theta(n-2) +\left[\frac{1}{4} n(n^3-6n^2+11n-6)\right]\Theta(n-4)  \label{freq2}
\\& \qquad -2(2+3n+n^2)(3+2n)^2 -\frac{1}{4} (2+3n+n^2)(n^2+7n+12)^2 \bigg \} \notag,
\end{align}
where $H_0|n\rangle =\omega_n^{(0)}|n\rangle$ defines the state $|n \rangle$ of the unperturbed harmonic oscillator, and $\Theta(n)=1$ for $n \geq 0$ and $\Theta(n)=0$ for $n<0$. The position translations described in the procedure can be implemented using the unitary translation operator,
\begin{align}
T(d)=\exp[-i \hat{p} d],
\end{align}
where $d$ is the displacement size, $d=d_1$ or $d=d_2$.  The matrix elements of this operator in the harmonic oscillator basis are
\begin{align}
\label{tmat}
\langle m|T(d)|n \rangle =\exp \left [- \frac{\gamma^2}{2} \right ] \sum_{q=0}^n \frac{(-\gamma)^q \gamma^{m-n+q} \sqrt{n!}\sqrt{ m!}}{(m-n+q)! (n-q)! q!} \Theta(m-n+q),
\end{align}
where $\gamma =d /\sqrt{2}$.
There are two special cases of Eq. (\ref{tmat}) that will be useful in our analysis.  The first is the displacement of the harmonic oscillator ground state $| n \rangle =| 0 \rangle$  (a Gaussian wavepacket centered at $x=0$),

\begin{align}
\label{t0}
\langle m|T(d)|0 \rangle =\exp \left [- \frac{\gamma^2}{2} \right ] \frac{\gamma^m}{\sqrt{m!}} \equiv \exp \left [- \frac{\gamma^2}{2} \right ]  C(m,\gamma),
\end{align}
where $C(m,\gamma)=\gamma^m/\sqrt{m!}$.  This expression is the well known coefficient for a coherent state formed by displacing the harmonic oscillator ground state.

The second useful expression is the case when $\gamma  \sqrt{n} \ll1$.  In this case, we can approximate the matrix element as

\begin{align}
\label{tsmall}
\langle m|T(d)|n \rangle & =\exp \left [- \frac{\gamma^2}{2} \right ] \Big [\delta_{m,n} +\delta_{m,n+1} \gamma \sqrt{n+1} -\delta_{m,n-1} \gamma \sqrt{n} \Theta(n-1)  \\
&\qquad -\delta_{m,n}  \gamma^2  n  \Theta(n-1)  +\frac{\delta_{m,n+2}}{2}  \gamma^2 \sqrt{(n+1)(n+2)} \notag \\
&\qquad+\frac{\delta_{m,n-2}}{2} \gamma^2\sqrt{n(n-1)} \Theta(n-2)  + \mathcal{O}(\gamma^3) \Big ] \notag
\end{align}
noting that small displacements predominantly excite nearby states.  

We first  consider the response of the system due to an initial displacement at $t=0$. Beginning with the harmonic oscillator ground state ($|\psi \rangle=|0\rangle$) at $t=0$, we apply a displacement $d_1$.  Using the described approximations for the frequencies, and, approximating the eigenstates as the unperturbed states $|n\rangle$, for a time $0<t<\tau$, the quantum state is
\begin{align}
|\psi (t<\tau) \rangle &= \exp \left [- \frac{\gamma_1^2}{2} \right ] \sum_n C(n,\gamma_1)  \exp [-i\omega_n t] |n \rangle .
\end{align}
After the displacement $d_1$, the  probability of being in the $n$-th state is  $\exp[-\gamma_1^2]C(n,\gamma_1)^2$, a Poisson distribution with mean,
\begin{align}
\bar{n}&=\gamma_1^2 ,\label{barn} 
\end{align}
and variance 
\begin{align}
\sigma_n^2&=\bar{n}=\gamma_1^2 \label{sigma}.
\end{align}
(Physically, the value of $\bar{n}$ is given as the classical energy over the harmonic oscillator frequency, $\omega_0=1$, $\bar{n}=(p^2/2+x^2/2)/\omega_0=d_1^2/2=\gamma_1^2$.) 

The response of the system, as measured by the expectation value,  $\langle \psi|x| \psi \rangle$, is given by
\begin{align}
\label{x0}
\frac{\langle x(t) \rangle}{2A^2/\sqrt{2}} =\sum_{n=0}^\infty C(n) C(n+1) \sqrt{n+1} \cos[(\omega_{n+1}-\omega_n)t],
\end{align}
where for notational clarity we  have dropped $\gamma_1$ in the argument of $C(n)=C(n,\gamma_1)$, and $A=\exp[-\gamma_1^2/2]$.

\begin{figure}[h!]
\begin{center}
\includegraphics*[height=0.45\textwidth,angle=0,clip]{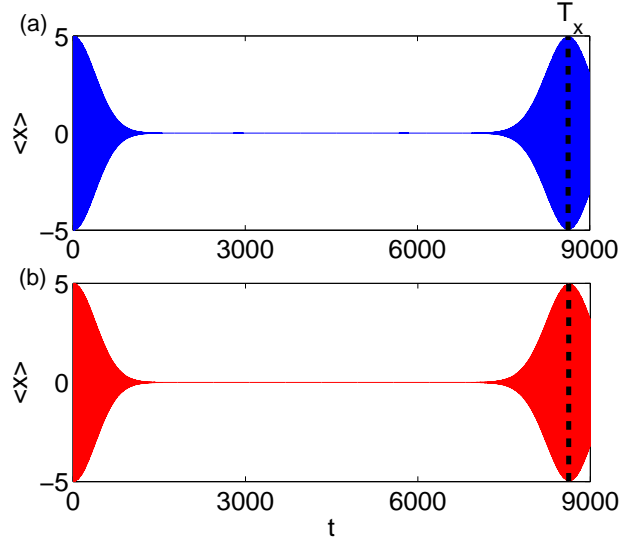}
\caption{ {\bf (a): } $\langle x(t) \rangle$ found through numerical solution of the Schr\"odinger Equation for a system with only one initial displacement at $t=0$,  $\beta=0.001$, $d_1$=5.  {\bf (b): } $\langle x(t) \rangle$ found through the perturbative model, for the same parameters as (a).  Note good agreement between the model and the numerical solution.
}
\label{b01_V5_V000}
\end{center}
\end{figure}

For comparison, we obtained numerical solutions of the Schr\"odinger Equation using a split-step operator method \cite{Gardiner2000}.
Figure \ref{b01_V5_V000}(a) shows a plot of $\langle x(t) \rangle$ versus $t$ obtained by numerical solution of the Schr\"odinger Equation in the absence of the second displacement. We see from Fig. \ref{b01_V5_V000}(a) that the displacement at $t=0$ leads to a fast oscillatory response whose envelope decays to small values by $t \sim 1500$.  At longer time, the response reassembles with its envelope reaching a peak value  at the `revival time' $T_x\cong8625$.  This revival of the response has been called a quantum revival or recurrence.    (Note that in Fig. \ref{b01_V5_V000} the fast oscillations are not visible because the thickness of the line that is plotted exceeds the period of the fast oscillations.)  Figure  \ref{b01_V5_V000}(b) plots the value of $\langle x(t) \rangle$ calculated from Eq. (\ref{x0}) with Eqs. (\ref{freqall}), (\ref{freq1}) and (\ref{freq2}) used for the frequencies.  We note very good agreement between the model and the numerical solution.

As we now verify, this term includes the well known quantum recurrence \cite{Raithel1998,Pitaevskii1997,Robinett2004}.  (The expression for Eq. (\ref{x0}) does not capture the fractional revival described in \cite{Kasperkovitz1995,Manfredi1996}.  This revival is much smaller in amplitude than the large revival at $t\approx T_x$, and is not visible on the scale used in Fig. \ref{b01_V5_V000}.  We have verified that a model including $\mathcal{O}(\beta)$  corrections to the energy eigenstates does reproduce these small fractional revivals). In order to estimate the quantum recurrence time of $\langle x(t) \rangle $, in a manner similar to Refs. \cite{Pitaevskii1997,Robinett2004}, we expand our expression for the frequency difference around a mean value of $n$, denoted by $\bar{n}$.  Using Eqs. (\ref{freq1}) and (\ref{freq2}), the energy difference between two adjacent states can be approximated as 
\begin{align}
\label{dw}
\omega_{n+1}-\omega_n \approx \Delta\omega_{\bar{n}} + a (n-\bar{n}) + ..., 
\end{align}
where 
\begin{align}
\Delta \omega_{\bar{n}}&=1+\frac{3}{4} \beta (\bar{n}+1) +\beta^2 \left (\frac{9}{8} +\frac{51}{32} \bar{n}  +\frac{51}{64} \bar{n}^2 \right ) \\ 
a&=\frac{3}{32}(8\beta -17\beta^2-17 \bar{n} \beta^2).
\end{align} 
$\Delta\omega_{\bar{n}}$ is the difference in the two energies, evaluated at $\bar{n}$ (and is independent of $n$). We can then express  $\cos[(\omega_{n+1}-\omega_n)t] $ as:
\begin{align}
\label{cosine}
\cos[(\Delta\omega_{\bar{n}}+a(n-\bar{n}))t] & \cong Re\big \{ \exp[i\Delta\omega_{\bar{n}}t]\exp[ia(n-\bar{n}))t] \big  \} .
\end{align}
In the vicinity of $\bar{n}$, the oscillatory behavior is a product of a fast oscillation at a frequency $\Delta \omega_{\bar{n}}$, which is independent of $n$, and a slow oscillation, which is periodic for all $n$ at a period $2\pi/a$.  Thus the envelopes  are periodic  at times $t=m T_x$ ($m=1,2,...$) where 
\begin{align}
\label{Tx}
T_x=\frac{2\pi}{a}, 
\end{align}
which agrees with the value $T_x\cong 8625$ estimated by inspection of Fig. \ref{b01_V5_V000}(a).  

It should also be noted that the time $T_x \approx  T_{rev}/2$, where $T_{rev}$ is the revival time it takes the state to come back to approximately its initial condition.  The recurrence in $\langle x(t)  \rangle$ corresponds to a mirror revival \cite{[][{.  This is a review article on quantum revival phenomena and contains additional references for the interested reader}]Robinett2004}, e.g.: the quantum state has reassembled as a mirror image of its initial condition on the opposite side of the anharmonic well.

Now, including the effect of the second displacement, the state is allowed to evolve to time $t=\tau$, at which time it experiences a small second  displacement by an amount $d_2$.   In order to utilize the expression in Eq. (\ref{tsmall}) we define $\gamma_2=d_2/\sqrt{2}$ and require  $\gamma_2 \sqrt{n} \ll 1$.  Taking as an upper bound on $n$ to be $\bar{n}+3\sigma_n$,  and noting that $\bar{n}= \gamma_1^2$ we require  $\gamma_2 \ll (\gamma_1^2+3\gamma_1)^{-1/2}$.
Using the expression for $\langle m|T(d)|n \rangle$ in Eq. (\ref{tsmall}) we find

\begin{align}
|\psi(t>\tau) \rangle & =\exp \left [- \frac{\gamma_1^2}{2} \right ] \sum_{m,n} \langle m|T(d_2)|n \rangle C(n) \exp[-i \omega_n \tau] \exp[-i \omega_m (t-\tau) ] | m \rangle  \notag \\
&\approx A \bigg \{ \sum_{m=0}^\infty C(m)\exp(-i\omega_m t) |m \rangle \label{psit} \notag  \\
& \quad + \gamma_2 \sum_{m=1}^\infty C(m-1) \sqrt{m} \exp[-i \omega_{m-1} \tau] \exp[-i \omega_m(t-\tau) ] |m \rangle \notag \\
& \quad- \gamma_2 \sum_{m=0}^\infty C(m+1) \sqrt{m+1} \exp[-i \omega_{m+1} \tau] \exp[-i \omega_m(t-\tau)] |m \rangle  \notag \\
&\quad  -\gamma_2^2 \sum_{m=1}^\infty C(m) m \exp[-i \omega_mt] |m \rangle \notag \\
& \quad + \frac{\gamma_2^2}{2}\sum_{m=2}^\infty C(m-2) \sqrt{m(m-1)} \exp[-i \omega_{m-2} \tau] \exp[-i \omega_m(t-\tau) ] |m \rangle \notag \\
& \quad+ \frac{\gamma_2^2 }{2}\sum_{m=0}^\infty C(m+2) \sqrt{(m+1)(m+2)} \exp[-i \omega_{m+2} \tau] \exp[-i \omega_m(t-\tau)] | m \rangle   +\mathcal{O}(\gamma_2^3)\bigg \}  ,
\end{align} 
where $A=\exp \left[-(\gamma_1^2 +\gamma_2^2)/2 \right]$.   Using the expression for the quantum state for $t>\tau$,  we can calculate the expectation value of the position $\langle x(t) \rangle=\langle \psi|x| \psi \rangle$ and order the terms in $\langle x(t) \rangle$ according to their dependence on the second displacement $\gamma_2$, 
\begin{align}
\label{xall}
\langle x(t) \rangle =\langle x(t) \rangle ^{(0)} +\langle x(t) \rangle ^{(1)} +\langle x(t) \rangle^{(2)} +... .
\end{align}
The $0th$ order term $\langle x(t) \rangle ^{(0)}$ is the result in the absence of the second displacement and is given by Eq. (\ref{x0}).  Calculating the next highest term in $\gamma_2$ in our sum, we find an expression for our first revival echo.  which can be separated into two parts, 
\begin{align}
\label{x1all}
\langle x (t)  \rangle ^{(1)} =\langle x  \rangle ^{(1)}_{1} +\langle x   \rangle ^{(1)} _{-1},
\end{align}
where 
\begin{align}
\label{x11}
\frac{1}{\gamma_2} \frac{\langle x \rangle^{(1)}_{1} }{A^2/\sqrt{2}}&= 2\sum_{n=0}^\infty (n+1)  \left \{C(n)^2-C(n+1)^2 \right \} \cos[(\omega_{n+1}-\omega_{n})(t-\tau)], \\ \notag
\end{align}
 and 
\begin{align}
\label{x1m1}
\frac{1}{\gamma_2} \frac{\langle x \rangle^{(1)}_{-1} }{A^2/\sqrt{2}}&= +2 \sum_{n=1}^\infty C(n+1)C(n-1)\sqrt{n(n+1)} \bigg \{ \cos \Big[(\omega_{n+1}-\omega_n) t+(\omega_{n}-\omega_{n-1})\tau \Big] \\ \notag 
& \qquad \qquad- \cos \Big[(\omega_{n-1}-\omega_n)t +(\omega_{n}-\omega_{n+1})\tau \Big ]\bigg \}.
\end{align}.

The superscript refers to the power of $\gamma_2$ (ignoring the common dependence in the normalization constant $A$), and the subscript indicates the time of the echo, in units of $\tau$, relative to $t=T_x$.  For example, the first sum $ \langle x \rangle^{(1)}_{1}$ has a cosine dependence similar to the quantum recurrence in Eq. (\ref{x0}), except $t$ has been replaced with $t-\tau$.  Thus, we expect a large amplitude in this second term at $t=\tau$ and $t \approx m T_x+\tau$ ($m \geq 1$), which we call a post-revival echo.  The second sum  $ \langle x \rangle^{(1)}_{-1}$ has a time dependence which again depends on the difference of adjacent frequencies.  Note however, that if we take the spacing between adjacent energy levels to be approximately independent of $n$,  $\omega_{n+1}-\omega_n \approx \omega_{n}-\omega_{n-1}$, then 
$\cos[(\omega_{n+1}-\omega_n) t+(\omega_{n}-\omega_{n-1})\tau] \approx \cos[(\omega_{n+1}-\omega_n) (t+\tau)] $.  Thus, we expect the second sum to contribute to a pre-revival echo at $t \approx m T_x-\tau$.  Note that our expression for $\langle x \rangle^{(1)}_{-1}$ only applies for $t>\tau$.  Thus the $m$ and $\tau$ values for pre-revival echoes are restricted by the requirement that $m T_x >2\tau$ (e.g., for $\tau <T_x/2$ we have that $m\geq 1$ is permissible).


Continuing on to $\langle x (t)\rangle^{(2)}$, we can again separate this term by the location of the dominant echoes,
\begin{align}
\label{x2}
\langle x(t) \rangle ^{(2)} &=\langle x \rangle ^{(2)}_0+\langle x\rangle ^{(2)}_{2} +\langle x\rangle ^{(2)}_{-2} ,
\end{align}
where expressions for $\langle x \rangle^{(2)}_{0}$, $\langle x \rangle^{(2)}_{2}$, and $\langle x \rangle^{(2)}_{-2}$ are given in Appendix \ref{Ap}.

With a similar argument for the time dependence of the first order terms, approximating the spacing between levels to be almost constant, each sum in $\langle x \rangle ^{(2)}_0$ produces a second order response at $t \approx T_x$.  Likewise, the  second term $\langle x \rangle ^{(2)}_{2}$ produces an echo at $t \approx 2 \tau$ and and a post-revival echo at $t \approx m T_x+2 \tau$, while pre-revival echoes from the last term $\langle x \rangle ^{(2)}_{-2}$  occur at  $t \approx mT_x-2 \tau$ ($m \geq 1$).

\begin{figure}[h!]
\begin{center}
\includegraphics*[height=0.80\textwidth,angle=0,clip]{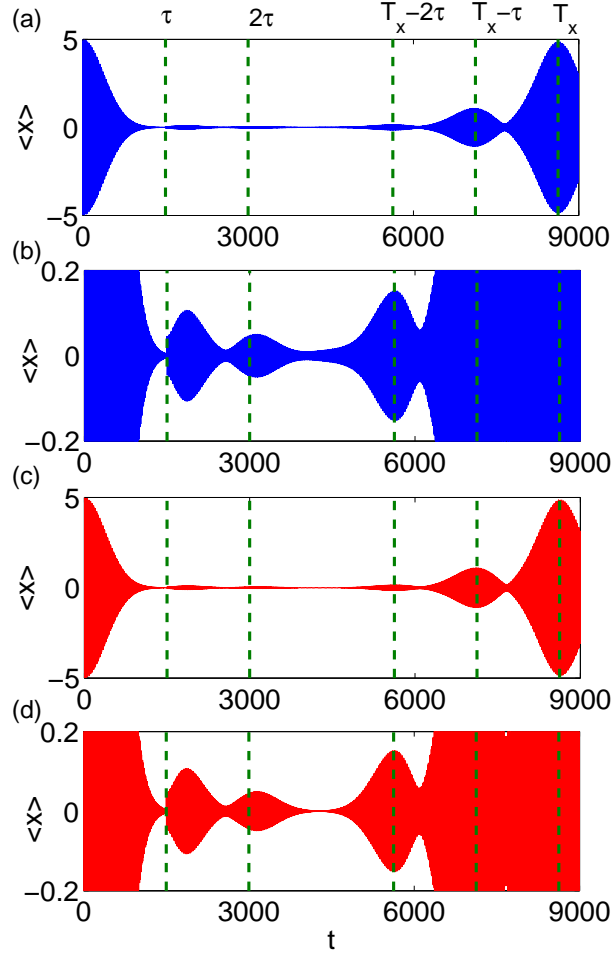}
\caption{ (Color Online): {\bf(a): } $\langle x(t) \rangle$ solved numerically for the full Schr\"odinger Equation for $\beta=0.001$, $d_1=5$, $d_2=0.05$, $\tau=1499$. {\bf(b):} same as (a), but zoomed in to display the smaller echoes.  {\bf (c):} The model $\langle x (t) \rangle $ plotted for the same parameters. {\bf (d):} same as (c), but zoomed in to display the smaller echoes.   We note that the model reconstructs the $t \approx T_x-\tau$ pre-revival echo, as well as the echo at $t \approx 2\tau$ and the pre-revival echo at $t \approx T_x-2\tau$. 
}
\label{b01_V5_V005_t1500}
\end{center}
\end{figure}

\begin{figure}[h!]
\begin{center}
\includegraphics*[height=0.50\textwidth,angle=0,clip]{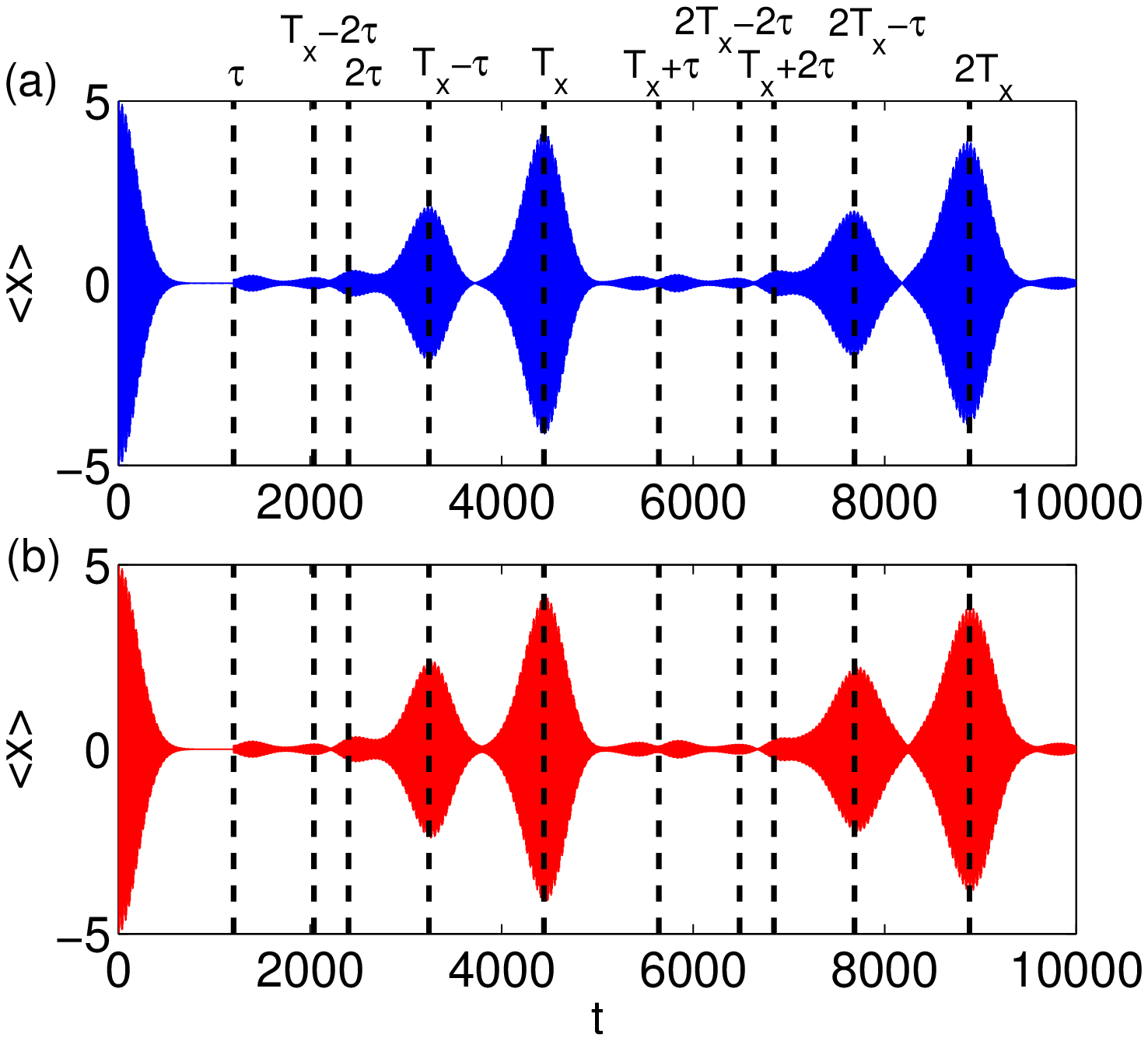}
\caption{ (Color Online):  {\bf(a)} $\langle x(t) \rangle$ solved numerically for the full Schr\"odinger Equation for $\beta=0.002$, $d_1=5$, $d_2=0.1$, $\tau=1200$.  {\bf (b)} The model $\langle x (t) \rangle $ plotted for the same parameters.  The model reconstructs the quantum recurrence at $t\approx m T_x$, as well as the echo and revival echo responses at $t\approx 2\tau$, $t \approx T_x +\tau$, $t\approx T_x-\tau$, and $t\approx T_x-2\tau$, $(m=1,2)$. 
}
\label{b02_V5_V01_t1200_zoom}
\end{center}
\end{figure}

Figures \ref{b01_V5_V005_t1500} and \ref{b02_V5_V01_t1200_zoom} compare numerical solutions of the Schr\"odinger Equation (Figs. \ref{b01_V5_V005_t1500}(a,b) and Fig. \ref{b02_V5_V01_t1200_zoom}(a)) with predictions (Figs. \ref{b01_V5_V005_t1500}(c,d) and Fig. \ref{b02_V5_V01_t1200_zoom}(b)) of the perturbation theory model [i.e., Eqs.  (\ref{x0}), (\ref{x1all})-(\ref{x1m1}), (\ref{x2})-(\ref{x2m2}) for $\langle x(t) \rangle$, with Eqs. (\ref{freqall}), (\ref{freq1}), and (\ref{freq2}) for $\omega_n$]  for different choices in $\beta$, $\tau$, and $d_2$ parameters.  In Fig. \ref{b01_V5_V005_t1500}, $\langle x (t) \rangle$ is calculated for $\beta=0.001$, $\tau=1499$, $d_2=0.05$.  There is good agreement between the numerics and the model, with the model reconstructing the quantum recurrence at $t\approx T_x$,  and the  echo and revival echoes at $t\approx 2\tau$, $t\approx T_x-\tau$  and $t \approx T_x-2\tau$. 

Similarly, Fig. \ref{b02_V5_V01_t1200_zoom} plots $\langle x(t) \rangle$ for a larger $\beta=0.002$ and $d_2=0.1$,  and a smaller value of $\tau=1200$,  showing more than one quantum revival echo.   There is again good agreement between the model and the numerics in capturing the behavior of the quantum recurrence at $t \approx m T_x$ ($m=1,2$) and the echo and revival echoes at $t \approx 2\tau$, $t \approx T_x +\tau$, $t \approx m T_x-\tau$ and $t \approx m T_x-2\tau$ ($m=1,2$).


\begin{figure}[h!]
\begin{center}
\includegraphics*[height=0.50\textwidth,angle=0,clip]{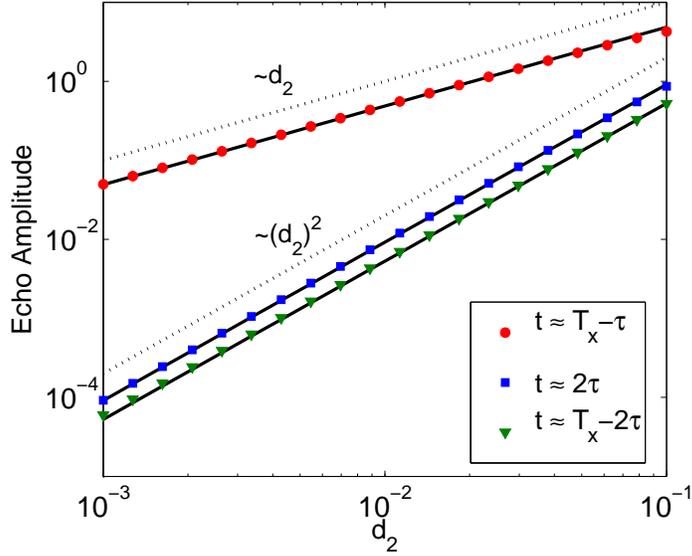}
\caption{(Color Online): The amplitude of three echoes present at different times as a function of $d_2$ for $d_1=5$, $\beta=0.001$, $\tau=1899$, ($t \approx T_x-\tau$: circles, $t \approx T_x-2\tau$: squares, $t \approx 2\tau$: triangles). The solid lines are echo response amplitudes extracted from the perturbation theory model. The dotted lines are guides to the eye indicating linear and quadratic behavior.  We note that the pre-revival echo at $t \approx T_x-\tau$ scales linearly with $d_2$, while the echo responses at $t \approx 2\tau$ and $t \approx T_x-2\tau$ scale quadratically. }
\label{anharm_scan_b01_V1_5}
\end{center}
\end{figure}

Additionally, the model indicates that one should expect the amplitude of the pre-revival echo a $t \approx T_x-\tau$ to scale linearly with the second displacement $d_2$.  As seen in Eq. (\ref{x11})  the term $\langle x \rangle ^{(1)}_1$ is proportional to $\gamma_2$ (and therefore $d_2$) if we ignore the weak dependence due to the normalization constant $A$.    Similarly, the echo at $t  \approx  2\tau$ and the pre-revival echo at $t \approx T_x-2\tau$ (represented by $\langle x\rangle^{(2)}_2$ and $\langle x\rangle^{(2)}_{-2}$) are expected to increase quadratically with $d_2$, as shown in Eqs. (\ref{x22}) and (\ref{x2m2})  of Appendix \ref{Ap}. We study the scaling of these echoes using the full numerical Schr\"odinger Equation with the choice of $\beta=0.001$,  $\tau=1899$, $d_1=5$, and various values of $d_2$.  This value of $\tau$ is chosen to avoid the coincidence of the $t  \approx  2\tau$ echo and the $t  \approx  T_x-2\tau$ pre-revival echo with the small fractional revival described in \cite{Kasperkovitz1995,Manfredi1996}.  The amplitudes of the echoes and pre-revival echoes as a function of $d_2$ are plotted in Fig. \ref{anharm_scan_b01_V1_5}, and compared to the amplitude scaling predictions of the perturbative model.  We note that indeed the pre-revival echo at $t  \approx T_x-\tau$ appears to behave linearly with $d_2$, while the echo at $t  \approx 2\tau$ and the pre-revival echo $t \approx  T_x-2 \tau$ appear to grow quadratically, as expected.  In general, the model suggests that echoes at $t \approx  n\tau$ and revival echoes at $t \approx  m T_x\pm n\tau$ ($m\geq 1$, $n \geq 1$) should have lowest order scalings of $(d_2)^n$.  (It should be noted, however, that the scaling of very small echoes may be affected by revivals and echoes occurring due to corrections in the energy eigenstates.)

\begin{figure}[h!]
\begin{center}
\includegraphics*[height=0.50\textwidth,angle=0,clip]{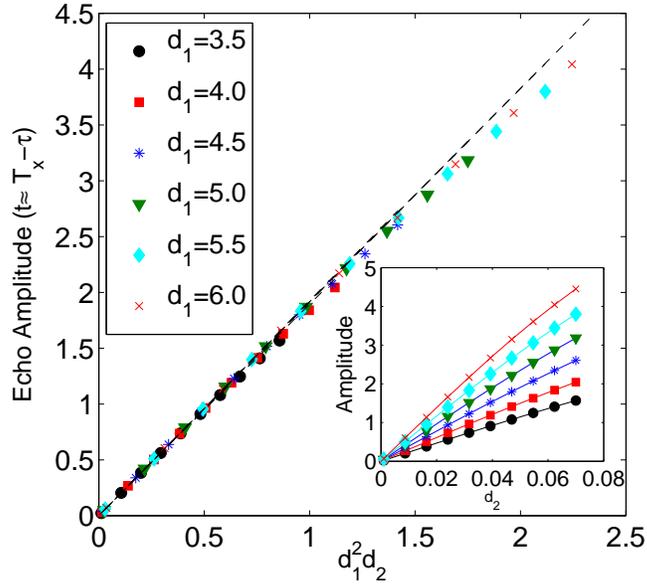}
\caption{(Color Online): 
The amplitude of the pre-revival echo at $t \approx T_x-\tau$ as a function of $d_1^2  d_2$ for $\beta=0.001$, $\tau=2499$, and various values of $d_1$ ($d_1=3.5$: circles, $d_1=4.0$: squares, $d_1=4.5$: asterisks, $d_1=5.0$: triangles, $d_1=5.5$: stars,   $d_1=6.0$: x's).  $d_2$ is allowed to vary between $0.001 \leq d_2 \leq 0.07$.  The dotted lines are echo amplitudes extracted from the model equations.  We note that the pre-revival echo at $t \approx T_x-\tau$  scales linearly with $d_1^2  d_2$. {\bf (Inset):}  The amplitude of the revival echo as a function of $d_2$, where lines link simulations done at equal values of $d_1$.
}
\label{scan_V1_V2_t2500}
\end{center}
\end{figure}

An approximate dependence of the pre-revival echo at $t \approx  T_x-\tau$ on the first perturbation $d_1$ can also be obtained, and can be understood by studying Eq. (\ref{x11}).
%
%
At the time of maximal $\langle x(t)  \rangle $ (in the vicinity of $t \approx  T_x-\tau$), one expects that for $n$ near $\bar{n}$, the cosine terms will be approximately in phase and only depend weakly on $n$.  Thus, taking the cosine terms to be roughly constant, and independent of $n$,
\begin{align}
\langle x \rangle^{(1)}_{-1} & \sim \gamma_2 \exp[-\gamma_1^2] \exp[-\gamma_2^2]\sum_{n=1}^\infty C(n+1)C(n-1)\sqrt{n(n+1)}
\end{align}
and using definitions of $C(n)$, $A$, and $\exp[\gamma_1^2]=\sum_{n=0}^\infty (\gamma_1^2)^n/n!$, we find

\begin{align}
\langle x \rangle^{(1)}_{-1} \sim  \gamma_1^2 \gamma_2 \exp[-\gamma_2^2] \sim  \gamma_1^2 \gamma_2 \sim d_1^2 d_2.
\end{align}
Note that we ignore any dependence on $d_1$ contained in the cosine terms (arising from  dependence on $\bar{n}$ of the expansion of $\omega_{n+1}-\omega_n$, which in turn is related to $d_1$).  Figure \ref{scan_V1_V2_t2500} plots the amplitude of the pre-revival echo as a function of $d_1^2  d_2$.  Note that echo amplitude appears to scale linearly with $d_1^2  d_2$.  For comparison, in the inset we plot the amplitude as a function of $d_2$, where lines connect simulations with equal values of $d_1$.

\begin{figure}[h!]
\begin{center}
\includegraphics*[height=0.50\textwidth,angle=0,clip]{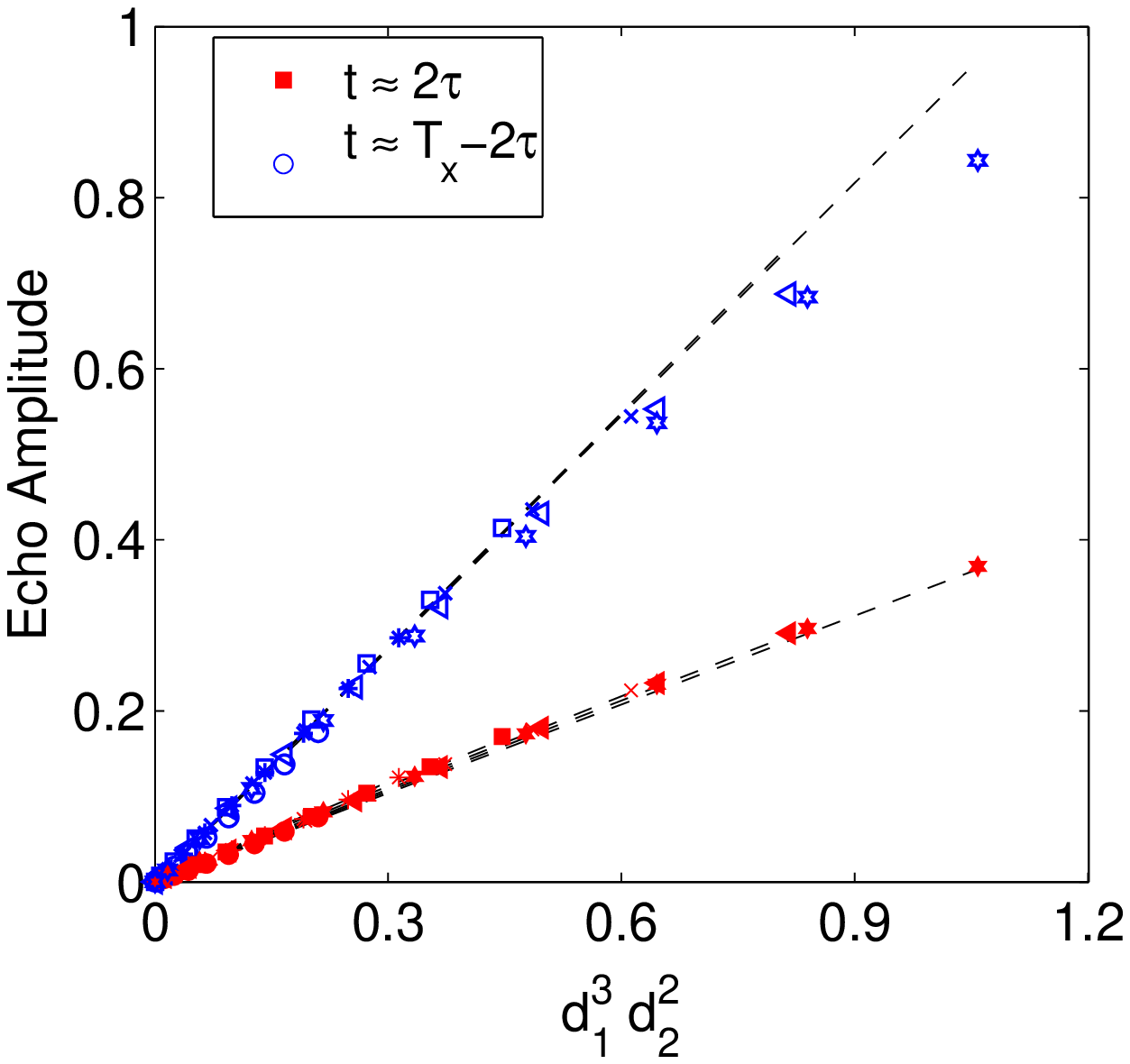}
\caption{(Color Online): 
The amplitude of the echo at $t \approx 2\tau$  (red, solid)  and the pre-revival echo at $t \approx T_x-2\tau$ (blue, unfilled) as a function of $d_1^3  d_2^2$ for, $\beta=0.001$, $\tau=1699$, and various values of $d_1$ and $d_2$.  $3.5 \leq d_1 \leq 6$ and $0.001 \leq d_2 \leq 0.07$.  The dotted lines are echo amplitudes extracted from the model equations.  We note that the $t\approx 2\tau$ echo and the pre-revival echo at $t \approx T_x-2\tau$ scale linearly with $d_1^3  d_2^2$.
}
\label{scan_V1_V2_1700}
\end{center}
\end{figure}

A similar procedure can be implemented for the response at $t \approx \tau$, and the echoes at  $t\approx T_x+\tau$, $t\approx 2\tau$ and  $t\approx T_x \pm 2\tau$.  The response at $t\approx \tau$ and the revival echo at $t\approx T_x+\tau$ are given by Eq. (\ref{x11}). Summing over the coefficient gives
\begin{align}
\langle x \rangle ^{(1)}_1 & \sim  d_2 
\end{align}
with approximately no $d_1$ dependence, which we numerically verify but do not show.  For the echo at $t\approx 2\tau$ and the revival echoes at $t\approx T_x \pm 2\tau$,  the oscillatory terms in Eqs.  (\ref{x22}) and (\ref{x2m2}) are assumed to be in phase, and the summations are done over the coefficients, leading to 
\begin{align}
\langle x \rangle ^{(2)}_2 & \sim d_1^3 d_2^2 + \mathcal{O}(d_1d_2^2) \\
\langle x \rangle^{(2)}_{-2} & \sim d_1^3 d_2^2 
\end{align}
Thus we expect the  echoes and revival echoes to approximately scale as $d_1^3 d_2^2$.  Figure \ref{scan_V1_V2_1700}, which plots the echo amplitudes as functions  of $d_1^3 d_2^2$, for $\beta=0.001$, $3.5 \leq d_1 \leq 6$ and $0.001 \leq d_2 \leq 0.07$ and $\tau=1699$, confirms this expectation.  (The post-revival echo at $t\approx T_x+2\tau$ is the recurrence of the echo at $t\approx 2\tau$.  It has an amplitude very similar to the $t\approx 2\tau$ echo and is not shown in Fig. \ref{scan_V1_V2_1700} for clarity.)

\begin{figure}[h!]
\begin{center}
\includegraphics*[height=0.8\textwidth,angle=0,clip]{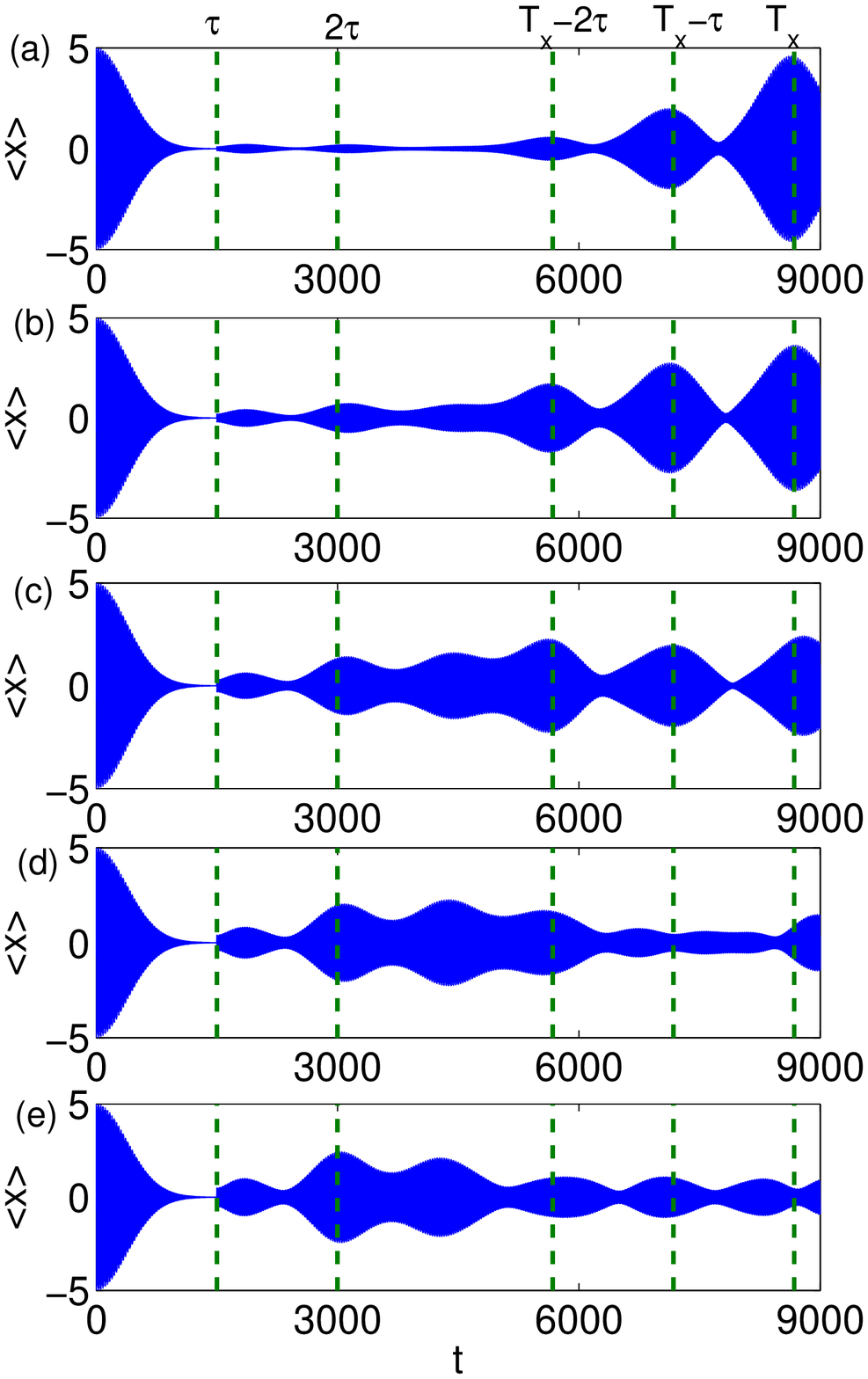}
\caption{(Color Online):  $\langle x(t) \rangle$ for $\beta=0.001$, $\tau=1499$, $d_1=5$, and different values of $d_2$,  {\bf (a) } $d_2=0.1$, {\bf (b) } $d_2=0.2$, {\bf (c) } $d_2=0.3$, {\bf (d) } $d_2=0.4$, and {\bf (e) } $d_2=0.5$.  Note that both the quantum recurrence at $t \approx T_x$ and the pre-revival echo at $t \approx T_x-\tau$ are suppressed as $d_2$ increases.  Further, the amplitude of $\langle x (t)  \rangle$ is larger for shorter times as $d_2$ increases.
}
\label{u0_increase_d2}
\end{center}
\end{figure}

Increasing $d_2$ to be $\sim 1$ (where we no longer expect our perturbative model to be valid), Fig. \ref{u0_increase_d2} plots the values of $\langle x(t) \rangle$ for $\beta=0.001$, $\tau=1499$, $d_1=5$, and different values of $d_2$ ranging from $d_2=0.1$ to $d_2=0.5$.  As $d_2$ is increased, we see that the amplitude of the quantum recurrence at $t \approx T_x$ diminishes.  Further, the amplitude of the pre-revival echo at $t \approx T_x-\tau$  is also suppressed, while the amplitude of $\langle x(t) \rangle$ increases for shorter times.

\begin{figure}[h!]
\begin{center}
\includegraphics*[height=0.5\textwidth,angle=0,clip]{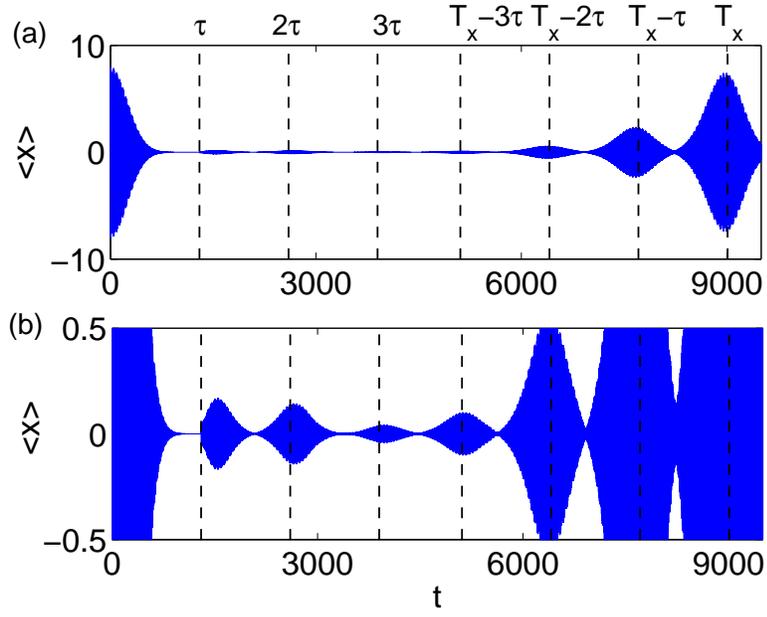}
\caption{(Color Online): {\bf (a) and (b): } $\langle x(t) \rangle$ sovled numerically for the full Schr\"odinger Equation for $\beta=0.001$, $d_1=8$, $d_2=0.05$ and $\tau=1299$.  Note that an echo at $t \approx 3\tau$ and pre-revival echo $t \approx T_x-3 \tau$ are clearly evident. (b) is the same as (a), plotted on a different scale to better display the echoes. }
\label{b01_V8_V05_t1300}
\end{center}
\end{figure}

Finally, we note that from numerical experiments solving the Schr\"odinger Equation, at larger $d_1$ we clearly see an echo at $t \approx 3\tau$ and revival echo at $t \approx T_x-3\tau$. An example illustrating this is shown in Fig. \ref{b01_V8_V05_t1300}.

\clearpage
\section{Dependence on the symmetry of the external perturbations}
\label{sym}
In Sec. \ref{u0dll1} we considered echoes and revival echoes induced by two successive displacements of a the trapping potential by amounts $d_1$ and $d_2$.  For $d_2$ small, we found that the revival echoes at $t\approx mT_x\pm \tau$ can be much larger (i.e., $\mathcal{O}(d_2)$ as opposed to $\mathcal{O}(d_2^2)$) than the first ordinary echo (i.e., the echo occurring at $t\approx 2 \tau)$.  A possibly significant point is that for small displacement $d$ and symmetric traps, $V(x)=V(-x)$, the perturbation may be viewed as being antisymmetric: $V(x+d)-V(x)= d (dV/dx) +\mathcal{O}(d^2)$.  Thus, a natural question is whether our result for the relative sizes of the $t \approx 2\tau$ echo and the revival echoes at $t \approx mT_x \pm \tau$ depend on the symmetry of the stimuli.  In what follows we consider this question and show that the answer is affirmative.

\begin{figure}[h!]
\begin{center}
\includegraphics*[height=0.4\textwidth,angle=0,clip]{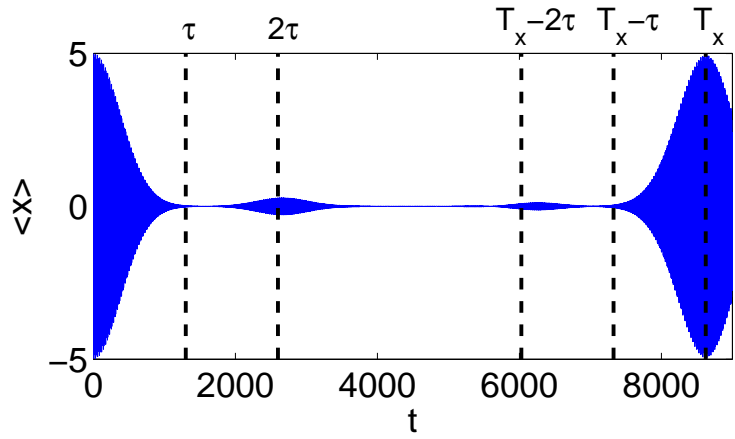}
\caption{(Color Online):  $\langle x(t) \rangle$ solved numerically for the full Schr\"odinger Equation with a displacement at t=0 ($d_1=5$), and an impulse squeeze at $t=\tau=1299$ ($\alpha_2=0.005$).  Notice there is no response at $t\tau$ or a pre-revival echo at $t\approx T_x-\tau$.  An echo at $t\approx 2\tau$ and a pre-revival echo at $t \approx T_x-2\tau$ are evident and are comparable in size. }
\label{V1_5__V2_0_005__squeeze_t1300}
\end{center}
\end{figure}

We have considered a different type of perturbation at $t=\tau$ which is symmetric in $x$, an impulse squeeze, i.e., we add a term $\alpha_2 x^2 \delta(t-\tau)$  to the potential.  
Taking $\beta=0.001$ and $\tau=1299$, we numerically calculate the response to an initial shift ($d_1=5$) and a small impulse squeeze ($\alpha_2=0.005$).  We display the results in Fig.  \ref{V1_5__V2_0_005__squeeze_t1300}.  In this case, echoes and revival echoes are evident in $\langle x (t) \rangle$, at $t \approx 2\tau$ and $t \approx T_x-2\tau$, with no responses in $\langle x(t)  \rangle$ at $t = \tau$ and $t \approx T_x-\tau$.  Model expressions for the echoes can be found in the same manner as in Sec. \ref{u0dll1}, and predict that the $t \approx 2\tau$ echo and $t \approx T_x-2\tau$ revival echo both scale linearly with $\alpha_2$ (see Appendix \ref{squeeze}).  Figure \ref{squeeze_V1_5__V2scan_t1300} plots the amplitude of the echo as a function of $\alpha_2$, demonstrating that the echo amplitudes depend linearly on $\alpha_2$.  This behavior is due to the impulse squeeze only exciting states of like parity; e.g., the  impulse squeeze acting on an even eigenstate only excites the other even eigenstates.  Note that unlike our result in Sec. \ref{u0dll1}, the first $\langle x(t)  \rangle$ echo (i.e., at $t \approx 2\tau$) and the first revival echo (i.e., that at $t \approx T_x-2\tau$) both scale linearly with the strength $\alpha_2$ of the second perturbation. (In fact it can be shown that all revival echoes at $t\approx mT_x\pm 2\tau$ scale linearly with $\alpha_2$.)

\begin{figure}[h!]
\begin{center}
\includegraphics*[height=0.5\textwidth,angle=0,clip]{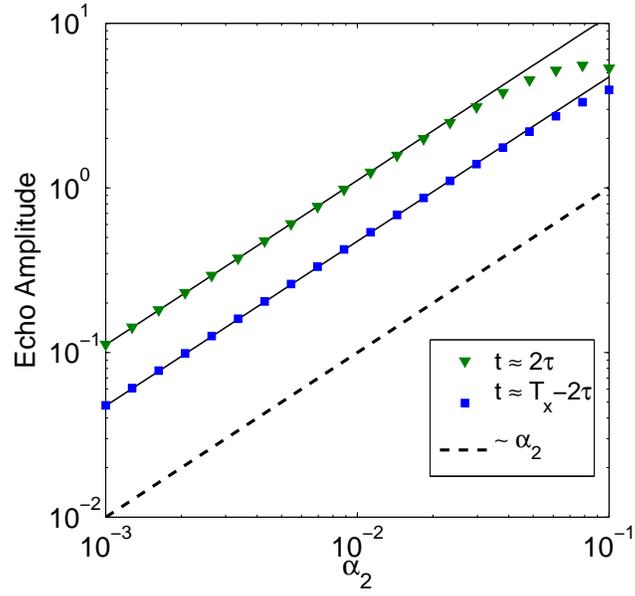}
\caption{(Color Online): The amplitude of the echo at $t \approx 2\tau$ (triangles) and the pre-revival echo at $t\approx T_x-2\tau$ (squares) as a function of $\alpha_2$ for, $d_1=5$, $\beta=0.001$, $\tau=1299$, and various values of $\alpha_2$.   The dotted line is a guide to the eye indicating linear behavior.  We note that both echoes scale linearly with $ \alpha_2$.  The solid lines are echo amplitudes extracted from a perturbation theory model (see Appendix \ref{squeeze}). }
\label{squeeze_V1_5__V2scan_t1300}
\end{center}
\end{figure}

\clearpage
\section{Quantum recurrences and revival echoes in $\langle x^p(t) \rangle$}
\label{xp}

Returning to the response to two successive displacements (as in Sec. \ref{u0dll1}), we now examine the behavior of the revival and the revival echoes in the response of $\langle x^p(t) \rangle$, for $p \geq 1$.  For example, if one were to measure the width of the cloud of cold atoms, the necessary observable would be  $w^2(t)= \langle x^2(t) \rangle - \langle x(t) \rangle^2 $.  Performing an analysis analogous to that of Sec. \ref{u0dll1} (described in detail in Appendix \ref{xpapp}), we find that quantum recurrences (in the absence of the second perturbation) are to be expected at fractional times of $t=T_x$.  In particular, for the observable $\langle x^p (t)\rangle$ , recurrences are expected at times 
$
t\approx mT_x/j
$
with $m \geq 1$.  The values of $j$ are contingent on whether $p$ is even or odd; for $p$ even: $j=2,4,\dots,p$, and for $p$ odd: $j=1,3,\dots,p$.    The amplitude of the recurrences is smaller for larger values of $j$. 

Additionally, revival echoes are also present in $\langle x^p (t) \rangle$.  The dominant echoes (those that scale as $d_2$), are also described by our model in Appendix \ref{xpapp}. Pre-revival echoes are expected at  $ t\approx (mT_x-\tau)/j$ and post-revivals at  $ t\approx (mT_x+\tau)/j$, where again $m \geq 1$ and the possible values of $j$ are different for even and odd $p$; for  $p$ even: $j=2,4,\dots,p$, and for $p$ odd: $j=1,3,\dots,p$.   For example, for $p=3$, recurrences are expected at $t \approx mT_x/3$, as well as larger revivals at $t \approx m T_x$,  $m\geq 1$.  Additionally there are dominant pre and post-revival echoes at $t\approx (mT_x \pm \tau)/3$ and $t\approx mT_x \pm \tau$, for $m\geq 1$ .   The model suggests that echoes at $t  \approx n \tau /j$, $(t \geq \tau)$, and revival echoes at $t  \approx (mT_x \pm n \tau)/j$ will scale as $\gamma_2^{n}$.  

\begin{figure}[h!]
\begin{center}
\includegraphics*[height=0.50\textwidth,angle=0,clip]{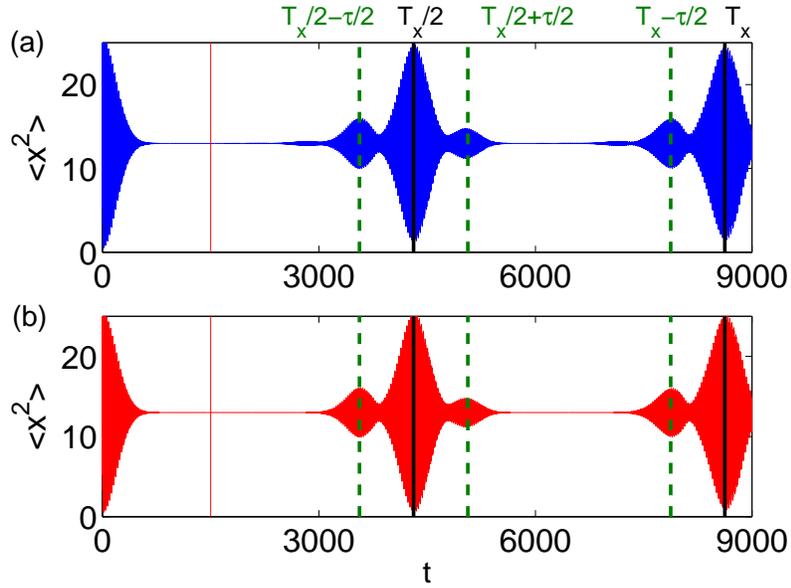}
\caption{(Color Online): {\bf (a) and (b): } $\langle x^2(t) \rangle$ sovled numerically for the full Schr\"odinger Equation for $\beta=0.001$, $d_1=5$, $d_2=0.05$ and $\tau=1499$.  Note that recurrences are present  at $t \approx T_x$ and $t \approx T_x/2$ (solid lines).  Also, pre-revival echoes at $t \approx (mT_x- \tau)/2$, $m=1,2$ and a post revival echo $t\approx (T_x+\tau)/2$ are evident (dotted lines). (b) The value of $\langle x^2(t) \rangle$ calculated using the model in Appendix \ref{xpapp}. }
\label{a1_b001_V1_5__V2_0p05_p2}
\end{center}
\end{figure}

\begin{figure}[h!]
\begin{center}
\includegraphics*[height=0.50\textwidth,angle=0,clip]{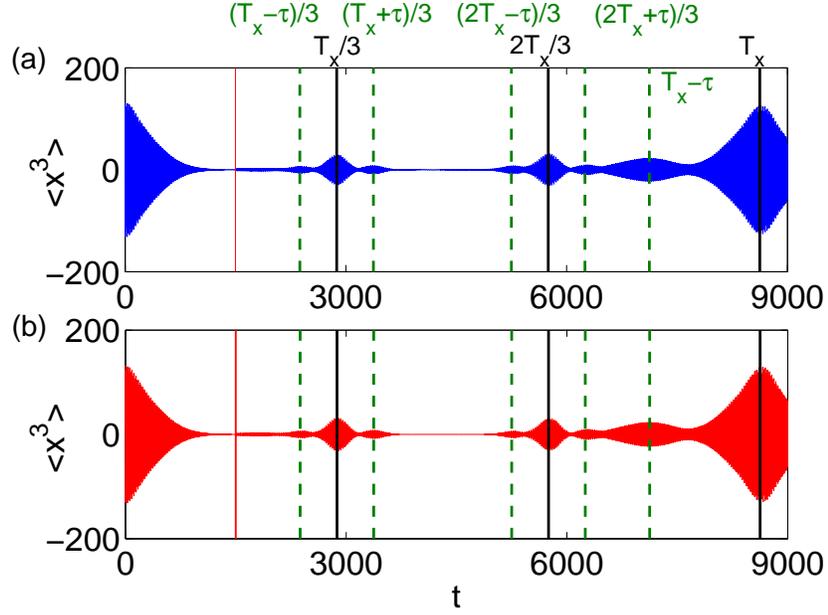}
\caption{(Color Online): {\bf (a) and (b): } $\langle x^3(t) \rangle$ sovled numerically for the full Schr\"odinger Equation for $\beta=0.001$, $d_1=5$, $d_2=0.05$ and $\tau=1499$.  Note that recurrences are present  at $t \approx T_x$ and $t \approx mT_x/3$, $m=1,2$ (solid lines).  Also, pre-revival echoes can be seen at $t \approx (mT_x- \tau)/3$, $m=1,2$ as well as $t\approx T_x- \tau$.  Post revival echoes are evident at $t\approx (mT_x+\tau)/3$ , $m=1,2$ (dotted lines). (b) The value of $\langle x^3(t) \rangle$ calculated using the model in Appendix \ref{xpapp}.}
\label{a1_b001_V1_5__V2_0p05_p3}
\end{center}
\end{figure}

Figures \ref{a1_b001_V1_5__V2_0p05_p2}(a)  and \ref {a1_b001_V1_5__V2_0p05_p3}(a) display the values of $\langle x^p(t) \rangle $ calculated by solving the Schr\"odinger Equation, for $p=2,$ and $p=3$ after an initial shift perturbation, $d_1=5$, at $t=0$, and a second shift perturbation, $d_2=0.05$, at $t=\tau=1499$, with $\beta=0.001$.  Displayed in solid lines are the expected times  of the quantum recurrences, and in dashed lines, the expected times of the pre and post-revival echoes.  For $p=2$, echoes at $t\approx 3\tau/2$ and $t\approx 4\tau/2$ are expected to be very small, scaling as $(d_2)^3$ and $(d_2)^4$, and are not readily evident upon examination of Figure \ref{a1_b001_V1_5__V2_0p05_p2}(a).  Regarding the scaling, as for the $p=1$ case, we note that the scaling of very small echoes may be affected by revivals and echoes occurring due to corrections in the energy eigenstates.

In Fig \ref{a1_b001_V1_5__V2_0p05_p3}, the pre-revival at $t\approx T_x-\tau/3$ is not visible because of the large quantum recurrence at $t\approx T_x$.  A conventional echo is expected at $t\approx 2\tau$, scaling as $(d_2)^2$ but is obscured by the revival at $t\approx T_x/3$.  For larger values of $d_2$, and shorter values of $\tau$, we have numerically observed the $t\approx 2\tau$ echo.  

Using the model expressions described in Appendix \ref{xpapp}, we compare the numerically calculated values of $\langle x^2(t) \rangle$ and  $\langle x^3(t) \rangle$ to those obtained from our model, as shown in Figs.  \ref{a1_b001_V1_5__V2_0p05_p2}(b)  and \ref {a1_b001_V1_5__V2_0p05_p3}(b).  Note that there is good agreement between the numerics and the model.  In both the $\langle  x^2(t)  \rangle$ and $\langle x^3(t)  \rangle$ cases, the response is dominated by the quantum recurrences, and the pre and post-\emph{revival} echoes.  

\clearpage
\section{Atom-atom interactions}
\label{u01_d1}

\begin{figure}[h!]
\begin{center}
\includegraphics*[height=0.9\textwidth,angle=0,clip]
{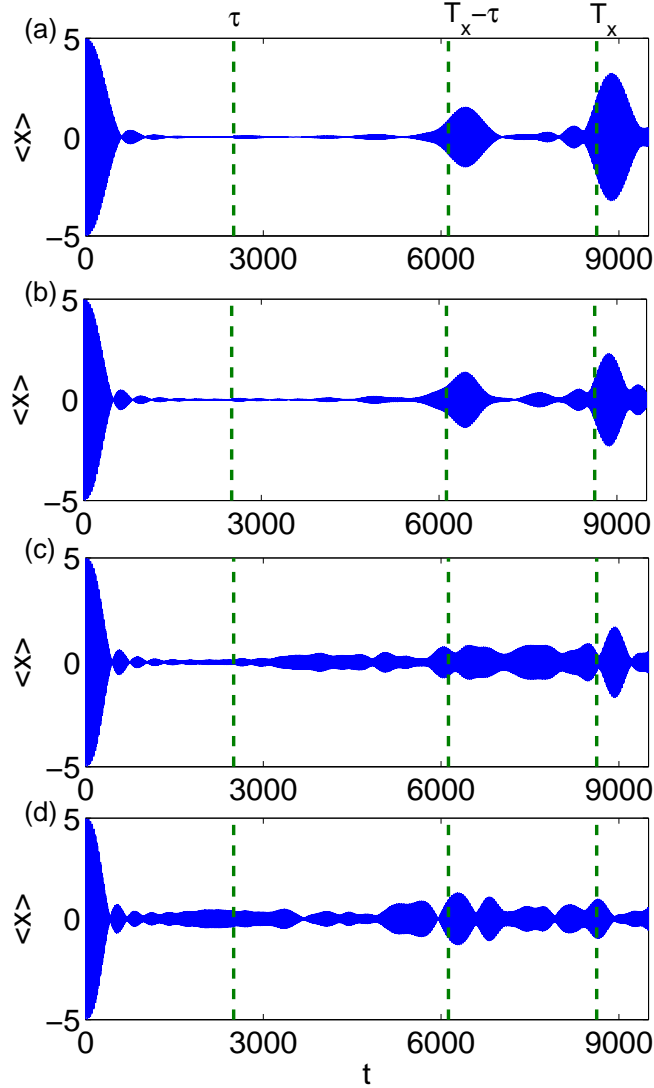}
\caption{(Color Online): The value of $\langle x(t) \rangle$ as a function of time with $d_1=5$, $d_2=0.05$, $\beta=0.001$, $\tau=2499$, for various values of $u$. {\bf (a)} $u=0.05$, {\bf (b)} $u=0.10$, {\bf (c)} $u=0.15$, {\bf (d)} $u=0.20$. We note that while increasing interactions causes damping to occur faster, it also leads to nonzero values of $\langle x(t) \rangle$ before the second kick at $t=\tau$.  Further, as $u$ is increased, both the quantum recurrence $t\approx T_x$ and the pre-revival echo at $t \approx T_x-\tau$ are suppressed. }
\label{udamp}
\end{center}
\end{figure}

Finally, we include a cubic nonlinear term in the Hamiltonian and numerically solve for the wavefunction. This can be considered as the mean field approximation for the condensate wavefunction of a Bose-Einstein condensate with interactions, and is known as the Gross-Pitaevskii Equation (GPE) \cite{Gross1961,Pitaevskii1961,Pethick2008}.

\ \begin{align}
i \frac{\partial}{\partial t } \psi & = (H_0 + H_1+H_{int}) \psi \\
H_{int} &=u|\psi|^2  \notag.
\end{align}
Here $u$ is a nonlinear interaction strength, which quantifies repulsive ($u>0$) or attractive ($u<0$) interactions in the condensate.  Again considering echo responses to two successive displacement (as in Sec. \ref{u0dll1}) and taking $d_1=5, d_2=0.05$, $\tau=2499$, and $\beta=0.001$ for various values of $u$, we numerically investigate the behavior of the echoes and revival echoes.  The results are plotted in Fig. \ref{udamp}.  As the interaction strength is increased, we note that the damping immediately after $t=0$ is increased, as compared to the $u=0$ case.  Additionally, for larger values of $u$, the value of $ \langle x(t) \rangle$ does not fully damp away before the application of the second displacement.  This observation has been made previously for similar systems \cite{Moulieras2012}.
The inclusion of interactions also suppresses the quantum recurrence at $t\sim T_x$, as well as the amplitude of the revival echo at $t\sim T_x-\tau$.  The time at which the quantum recurrence occurs is also shifted from the predicted value of $t\approx T_x$ in the noninteracting case.

\clearpage
\section{Summary and Conclusions}
In this paper, we study echoes and revival echoes for a collection of cold atoms in a weakly anharmonic potential subjected to two external stimuli, one at $t=0$ and the other at $t=\tau$.  In the case where  the two external stimuli are sudden displacements of sizes $d_1$ at $t=0$ and $d_2$ at  $t=\tau$, we observe responses in the expected value of position $\langle x(t)  \rangle$, and find that responses occur not only at $t \approx n\tau$, ($n\geq 1$) and at well-known quantum recurrences at $t\approx mT_x$, but also at $t \approx m T_x-n\tau > \tau$ (pre-revival echoes) and $t\approx mT_x+n\tau$ (post-revival echoes).  Again, in the case where the external stimuli are sudden displacements,  we have formulated a perturbation theory model and note good agreement between our model and the numerical results.  We numerically verify that for small $d_2$, the revival echo at $t \approx T_x \pm \tau$ scale linearly with $d_2$ while echoes at $t \approx 2\tau$ and revival echoes at $t \approx T_x \pm 2\tau$ scale as $ (d_2)^2$.  As expected from our perturbative model, and seen numerically, this scaling suggests that, for a sufficiently small $d_2$, the revival echoes at $t \approx T_x \pm \tau$ can be substantially  larger than the echo at $t \approx 2\tau$.  In addition, our perturbation theory model shows that for small $d_1$ and $d_2$, the revival echo at $t \approx T_x - \tau$  scales as $ d_1^2 d_2$, (which we numerically verify), while the echo at $t\approx 2\tau$ and the revival echoes at $t \approx T_x \pm 2\tau$ scale  as $ d_1^3 d_2^2$ (which we numerically verify).


We numerically demonstrate the suppression of the quantum revival $t\approx T_x$ and the revival echo at $t \approx T_x-\tau$, when the size of the second displacement is increased, $d_2 \sim 1$.  We also investigate how the echoes and revival echoes depend on the form and symmetry of the external stimuli.  One result is that, in the case where the first stimuli is a displacement,  the domination of the size of the revival echoes at $t\approx m T_x \pm \tau$ over the echoes at $t\approx 2\tau$ does not occur when the small second external stimulus has  different symmetry than that of a small displacement.  

The presence of quantum recurrences and revival echoes in responses observed in higher moments of position $x^p$ was also studied.  Quantum recurrences of $\langle x^p (t)\rangle$ appear at fractions of the $p=1$ revival time, $t \approx m T_x/j$, where $m =1,2,\dots$, and the value of $j$ is determined by $p$.  Revival echoes are also present in $x^p$, with dominant echoes (scaling as $d_2$) at $\langle x^p (t)\rangle$ at times $t\approx (mT_x\pm\tau)/j$.   

Finally, the the inclusion of interactions, modeled with the Gross-Pitaevskii equation, demonstrates that the quantum recurrence and the revival 
echoes  are suppressed as the nonlinear interaction strength $u$, is increased.

Imaging is the natural way to resolve revivals of $x$.  Revivals and echoes  in momentum, $p$, may be found by a similar calculation.  In this case, these revivals and echoes may be observed  if the trap is switched off at a revival or echo time.  The cloud of atoms will move together since all atoms have nearly the same momentum.  If instead, the trap is turned off at a time not coinciding with a revival or echo, the cloud of atoms will simply spread.
%
%
\section{Acknowledgements}
We thank Steve Rolston for his very useful comments.  This work was partly supported by the US/Israel Binational Science Foundation.
MH was supported by the Department of Defense (DoD) through the National Defense Science \& Engineering Graduate Fellowship (NDSEG) Program.
\appendix
\section{Expressions for $\langle x \rangle^{(2)}_{0}$, $\langle x \rangle^{(2)}_{2}$, and $\langle x \rangle^{(2)}_{-2}$ }
\label{Ap}
Continuing the procedure outlined in Sec. \ref{u0dll1}, we can find the terms in the expectation value of $\langle x(t) \rangle$ which are quadratic in $\gamma_2$.   These can again be separated into the location of their dominant echoes,
\begin{align}
\label{x2}
\langle x(t) \rangle ^{(2)} &=\langle x \rangle ^{(2)}_0+\langle x\rangle ^{(2)}_{2} +\langle x\rangle ^{(2)}_{-2} .
\end{align}
The expressions for for $\langle x \rangle^{(2)}_{0}$, $\langle x \rangle^{(2)}_{2}$, and $\langle x \rangle^{(2)}_{-2}$ are found to be:
\begin{align}
\label{x20}
\frac{1}{\gamma_2^2} \frac{\langle x \rangle ^{(2)}_0}{A^2/\sqrt{2}} &= 2\sum_{n=0}^\infty C(n+1)C(n+2) (n+1)\sqrt{n+2} \cos \Big[(\omega_{n+1}-\omega_{n})(t-\tau)+(\omega_{n+2}-\omega_{n+1}) \tau \Big ] \\ \notag
& \quad +2\sum_{n=1}^\infty C(n-1)C(n) (n+1)\sqrt{n}   \cos \Big [ (\omega_{n+1}-\omega_n)(t-\tau) +(\omega_n-\omega_{n-1})\tau \Big ] \\ \notag
& \quad -2 \sum _{n=0} C(n)C(n+1) (n+1)^{3/2}  \cos \Big [ (\omega_{n+1}-\omega_n)t\Big ] \\ \notag
& \quad -2 \sum _{n=2} C(n)C(n-1) \sqrt{n}(n-1)  \cos \Big [ (\omega_{n}-\omega_{n-1})t \Big ] \notag,
\end{align} 
 
\begin{align}
\label{x22}
\frac{1}{\gamma_2^2} \frac{\langle x \rangle ^{(2)}_2}{A^2/\sqrt{2}} &= -2\sum _{n=0}  C(n)C(n+1) (n+1)^{3/2}  \cos \Big [ (\omega_{n+1}-\omega_n)(t-\tau) +(\omega_n-\omega_{n+1})\tau \Big ]  \\
&  \quad +\sum_{n=1}^\infty C(n)C(n-1) (n+1)\sqrt{n}  \cos \Big[ (\omega_{n+1}-\omega_{n})t +(\omega_{n-1}-\omega_{n+1})\tau \Big ] \notag \\
&  \quad +\sum_{n=1}^\infty C(n)C(n+1) n\sqrt{n+1}  \cos \Big[ (\omega_{n}-\omega_{n-1})t +(\omega_{n-1}-\omega_{n+1})\tau \Big ] \notag,
\end{align} 
and
\begin{align}
\label{x2m2}
\frac{1}{\gamma_2^2} \frac{\langle x \rangle ^{(2)}_{-2}}{A^2/\sqrt{2}} &=  - 2 \sum_{n=2}^\infty  C(n-2)C(n+1) \sqrt{n(n^2-1)} \cos \Big [ (\omega_{n}-\omega_{n-1})(t-\tau) +(\omega_{n+1}-\omega_{n-2})\tau \Big ]  \\
&  \quad +\sum_{n=3}^\infty C(n)C(n-3) \sqrt{(n)(n-1)(n-2)}  \cos \Big[ (\omega_{n}-\omega_{n-1})t +(\omega_{n-1}-\omega_{n-3})\tau \Big ] \notag \\
&  \quad +\sum_{n=0}^\infty C(n)C(n+3) \sqrt{(n+1)(n+2)(n+3)}  \cos \Big[ (\omega_{n+1}-\omega_{n})t +(\omega_{n+3}-\omega_{n+1})\tau \Big ] \notag.
\end{align} 

In order to get the approximate scaling for the echo at $t \approx 2\tau$ and the revival echo at $t\approx T_x-\tau$, we take the above expressions for the echoes, and take the cosine terms to be in phase and independent of $n$.  Doing the summations for the coefficients, we find for $\langle x \rangle^{(2)}_{2}$:
\begin{align}
\sum_{n=0}^\infty C(n)C(n+1)(n+1)^{3/2} &= \exp[\gamma_1^2] (\gamma_1^3+\gamma_1) \\
\sum_{n=1}^\infty C(n)C(n-1)(n+1)\sqrt{n} &= \exp[\gamma_1^2] (\gamma_1^3+2\gamma_1) \\
\sum_{n=1}^\infty C(n)C(n+1)n\sqrt{n+1} &= \exp[\gamma_1^2] (\gamma_1^3) ,
\end{align}
and for $\langle x \rangle^{(2)}_{-2}$:
\begin{align}
\sum_{n=2}^\infty C(n-2)C(n+1)]\sqrt{n(n^2-1)} &= \exp[\gamma_1^2] (\gamma_1^3) \\
\sum_{n=3}^\infty C(n)C(n-3)\sqrt{n(n-1)(n-2)} &= \exp[\gamma_1^2] (\gamma_1^3) \\
\sum_{n=0}^\infty C(n)C(n+3)n\sqrt{(n+1)(n+2)(n+3)} &= \exp[\gamma_1^2] (\gamma_1^3) .
\end{align}
The factor $\exp[\gamma_1^2]$ is canceled  by the $\gamma_1$ dependence of $A^2$, leading to the $d_1$ dependence described in Sec. \ref{u0dll1}.

\section{Model Expressions for an initial displacement, followed by an impulse squeeze.}
\label{squeeze}
In a manner analogous to the calculation presented in Sec. \ref{u0dll1}, The response of  $\langle x(t) \rangle$ to an initial displacement at $t=0$ ($d=d_1$) and an impulse squeeze at time $\tau$ $(\alpha=\alpha_2$ , $\alpha_2 (\bar{n}+3\sigma_n) \ll 1)$ can also be estimated.  Approximating the matrix elements of the squeeze operator in the unperturbed anharmonic oscillator basis, we find
\begin{align}
\langle m|\exp[-i\alpha_2 x^2 |n \rangle &\approx \delta_{m,n}-i\frac{\alpha_2}{2} \bigg \{ (2n+1)\delta_{m,n} +\sqrt{n(n-1)}\delta_{m,n-2} \\
&\quad +\sqrt{(n+1)(n+2)} \delta_{m,n-2} \bigg \} \notag.
\end{align}
From this, the responses proportional to $\alpha_2$ can be calculated,

\begin{align}
\langle x \rangle^{(1)} = \langle x \rangle^{(1)}_{2}+\langle x \rangle^{(1)}_{-2}
\end{align}
where
\begin{align}
\frac{\langle x \rangle^{(1)}_{2} }{\exp[-\gamma_1^2] \alpha_2/\sqrt{2}}&= \sum_{n=1}^\infty C(n)C(n+1)n\sqrt{n+1}\sin[(\omega_n-\omega_{n-1})t+(\omega_{n-1}-\omega_{n+1})\tau] \\
&+\sum_{n=1}^\infty C(n)C(n-1)(n+1)\sqrt{n}\sin[(\omega_n-\omega_{n+1})t+(\omega_{n+1}-\omega_{n-1})\tau] \notag
\end{align}
and 
\begin{align}
\frac{\langle x \rangle^{(1)}_{-2} }{\exp[-\gamma_1^2] \alpha_2/\sqrt{2}}&= \sum_{n=0}^\infty C(n)C(n+3)\sqrt{(n+1)(n+2)(n+3)} \\
& \qquad \times \sin[(\omega_n-\omega_{n+1})t+(\omega_{n+1}-\omega_{n+3})\tau] \notag  \\
&+\sum_{n=3}^\infty C(n)C(n-3)\sqrt{n(n-1)(n-2)}  \notag \\
& \qquad \times \sin[(\omega_n-\omega_{n-1})t+(\omega_{n-1}-\omega_{n-3})\tau]  \notag.
\end{align}

\section{Model Expressions for $\langle x^p(t) \rangle$ after two shift perturbations.}
\label{xpapp}
In the case of a shift perturbation, $d_1$, at $t=0$, and a smaller shift perturbation, $d_2$, at time $t=\tau$, we derive model equations for both the quantum recurrence, as well as revival echoes in the quantity 
\begin{align}
\langle x^p(t) \rangle= \langle \psi|x^p|\psi \rangle
\end{align}
for $p\geq1$.  Using the definition of $x$ in Eq. \ref{xdef}, one can write the matrix elements of $x^p$, in the unperturbed harmonic oscillator basis, as

\begin{align}
\langle m|x^p|n \rangle &=\left( \frac{1}{\sqrt{2}} \right )^p \langle m|(\hat{a}+\hat{a}^{\dagger})^p|n \rangle \\
&=\left( \frac{1}{\sqrt{2}} \right )^p  \sum_{j=0}^p B(n,p-2j,p) \delta_{m,n+p-2j} \label{xpelem}
\end{align}

where  $B(n,p-2j,p)$ are coefficients that can be found via application of the annihilation and creation operators.  For example, for $p=3$,  the coefficients are 

\begin{align}
B(n,-3,3)&=\sqrt{n(n-1)(n-2)} \\
B(n,-1,3)&=3n\sqrt{n} \\
B(n,1,3)&=3(n+1)\sqrt{n+1} \\
B(n,3,3)&=\sqrt{(n+1)(n+2)(n+3)} .
\end{align}

It can be shown that these coefficients obey the relation $B(n,-d,p)=B(n-d,d,p)$.  Using Eq. (\ref{xpelem}), and the expression for the quantum state after two displacements, Eq. (\ref{psit}), one can again find the time dependence of the expectation value of $x^p(t)$, and order the terms according to their dependence on $\gamma_2$,
\begin{align}
\langle x^p(t) \rangle =  \langle x^p(t) \rangle^{(0)} + \langle x^p(t) \rangle^{(1)} + \dots
\end{align}
The term $\langle  x^p(t) \rangle^{(0)} $is again the quantum recurrence in the absence of the second displacement, and if $p$ is even, is given by,
\begin{align}
\label{recurr_even}
\frac{  \langle x^p(t) \rangle^{(0)}}{2A^2 (1/\sqrt{2})^p}= \sum_ {\substack{j=0\\ j \text{ even}}}^p \sum_{n=0}^\infty \left(1-\frac{\delta_{0,j}}{2} \right ) C(n)C(n+j) \cos[(\omega_{n+j}-\omega_n)t ] B(n,j,p)
\end{align}
and if $p$ is odd, 
\begin{align}
\label{recurr_odd}
\frac{  \langle x^p(t) \rangle^{(0)}}{2A^2 (1/\sqrt{2})^p}= \sum_ {\substack{j=1\\ j \text{ odd}}}^p \sum_{n=0}^\infty C(n)C(n+j) \cos[(\omega_{n+j}-\omega_n)t ] B(n,j,p) .
\end{align}
Employing the same argument for the behavior of the quantum recurrence in Sec. \ref{u0dll1}, one finds 
\begin{align}
(\omega_{n+j}-\omega_n )&= (\omega_{n+j}-\omega_{n+j-1}) +( \omega_{n+j-1}-\omega_{n+j-2}) + \dots+( \omega_{n+1}-\omega_n) \\
&\approx j\Delta_{\bar{n}}+ja(n-\bar{n}+q) \notag .
\end{align}
where $q$ is a constant independent of $n$.  Compared to the $j=1$ case of Sec. \ref{u0dll1}, the slow envelope maxima are now periodic for all $n$ at $2\pi/(ja)$.  Thus, each $\cos[(\omega_{n+j}-\omega_n)t]$ will give rise to a recurrence at time 
\begin{align}
t\approx \frac{mT_x}{j}, \quad (j \geq 1, m \geq 1)
\end{align}  
If $p$ is even, then the $j=0$ case is a constant which increases the value of $\langle x^p(t)  \rangle$.
Note that  the expectation value of $\langle x^p \rangle$ contains recurrences not only at $t\approx m T_x/p$, but also at the values of $j < p$ as seen in Eqs. (\ref{recurr_even}) and (\ref{recurr_odd}).

The expression for the revival echoes that are approximately linear to $d_2$ can also be found, for $p$ even:
\begin{align}
\label{xp1even}
\frac{\langle x^p(t) \rangle^{(1)}}{2 \gamma_2 (1/\sqrt{2})^p A^2}= \langle x^p \rangle^{(1)}_0 + \sum_{\substack{j=2\\ j \text{ even}}}^p \left \{ \langle x^p(t) \rangle^{(1)}_{j+} + \langle x^p(t) \rangle^{(1)}_{j-} \right \}
\end{align}
and $p$ odd:
\begin{align}
\label{xp1odd}
\frac{\langle x^p(t) \rangle^{(1)}}{2 \gamma_2 (1/\sqrt{2})^p A^2}= \sum_{\substack{j=1\\ j \text{ odd}}}^p \left \{ \langle x^p(t) \rangle^{(1)}_{j+} + \langle x^p(t) \rangle^{(1)}_{j-} \right \}
\end{align}
where,
\begin{align}
\langle x^p \rangle^{(1)}_0 = \sum_{n=0}^\infty C(n+1)C(n) \sqrt{n+1} \cos[(\omega_{n+1}-\omega_n)\tau] [B(n+1,0,p)-B(n,0,p)],
\end{align}

\begin{align}
\langle x^p(t) \rangle^{(1)}_{j+}& =\sum_{n=j}^\infty C(n-1)C(n-j)\sqrt{n} B(n,-j,p) \cos[(\omega_{n-j}-\omega_n)t+(\omega_n-\omega_{n-1})\tau]   \\
&\quad -\sum_{n=0}^\infty C(n+1)C(n+j)\sqrt{n+1} B(n,j,p) \cos[(\omega_{n+j}-\omega_n)t+(\omega_n-\omega_{n+1})\tau] \notag,
\end{align}
and
\begin{align}
\langle x^p(t) \rangle^{(1)}_{j-}& =\sum_{n=1}^\infty C(n-1)C(n+j)\sqrt{n} B(n,j,p) \cos[(\omega_{n+j}-\omega_n)t+(\omega_n-\omega_{n-1})\tau]   \\
&\quad -\sum_{n=j}^\infty C(n+1)C(n-j)\sqrt{n+1} B(n,-j,p) \cos[(\omega_{n-j}-\omega_m)t+(\omega_n-\omega_{n+1})\tau] \notag .
\end{align}

The term $\langle x^p \rangle^{(1)}_0$ is a time independent offset.  The quantity $ \langle x^p(t) \rangle^{(1)}_{j+} $ represents a  post-revival echo at time $t\approx (mTx+\tau)/j$.  Similarly the term $ \langle x^p(t) \rangle^{(1)}_{j-} $ gives rise to a pre-revival echo at  $t\approx(mT_x-\tau)/j$ , $j \geq 1$ and $m\geq 1$.  As can be seen from Eqs. (\ref{xp1even}) and (\ref{xp1odd}), revival echoes will be present and at various fractional shifts of $\tau$ from the different recurrences.   For example, for $p=3$, recurrences are expected at $t \approx \frac{mT_x}{3}$, as well as larger recurrences at $t \approx m T_x$,  $m\geq 1$.  Additionally there are dominant pre and post-revival echoes at $t\approx (mT_x \pm \tau)/3$ and $t\approx mT_x \pm \tau$, for $m\geq 1$ (see Figure \ref{a1_b001_V1_5__V2_0p05_p3}).


\begin{thebibliography}{10}%
\makeatletter
\providecommand \@ifxundefined [1]{%
 \ifx #1\undefined \expandafter \@firstoftwo
 \else \expandafter \@secondoftwo
\fi
}%
\providecommand \@ifnum [1]{%
 \ifnum #1\expandafter \@firstoftwo
 \else \expandafter \@secondoftwo
\fi
}%
\providecommand \enquote [1]{``#1''}%
\providecommand \bibnamefont  [1]{#1}%
\providecommand \bibfnamefont [1]{#1}%
\providecommand \citenamefont [1]{#1}%
\providecommand\href[0]{\@sanitize\@href}%
\providecommand\@href[1]{\endgroup\@@startlink{#1}\endgroup\@@href}%
\providecommand\@@href[1]{#1\@@endlink}%
\providecommand \@sanitize [0]{\begingroup\catcode`\&12\catcode`\#12\relax}%
\@ifxundefined \pdfoutput {\@firstoftwo}{%
 \@ifnum{\z@=\pdfoutput}{\@firstoftwo}{\@secondoftwo}%
}{%
 \providecommand\@@startlink[1]{\leavevmode}%
 \providecommand\@@endlink[0]{}%
}{%
 \providecommand\@@startlink[1]{%
  \leavevmode
  \pdfstartlink
   attr{/Border[0 0 1 ]/H/I/C[0 1 1]}%
   user{/Subtype/Link/A<</Type/Action/S/URI/URI(#1)>>}%
  \relax
 }%
 \providecommand\@@endlink[0]{\pdfendlink}%
}%
\providecommand \url  [0]{\begingroup\@sanitize \@url }%
\providecommand \@url [1]{\endgroup\@href {#1}{\urlprefix}}%
\providecommand \urlprefix [0]{URL }%
\providecommand \Eprint[0]{\href }%
\@ifxundefined \urlstyle {%
  \providecommand \doi [1]{doi:\discretionary{}{}{}#1}%
}{%
  \providecommand \doi [0]{doi:\discretionary{}{}{}\begingroup
  \urlstyle{rm}\Url }%
}%
\providecommand \doibase [0]{http://dx.doi.org/}%
\providecommand \Doi[1]{\href{\doibase#1}}%
\providecommand \bibAnnote [3]{%
  \BibitemShut{#1}%
  \begin{quotation}\noindent
    \textsc{Key:}\ #2\\\textsc{Annotation:}\ #3%
  \end{quotation}%
}%
\providecommand \bibAnnoteFile [2]{%
  \IfFileExists{#2}{\bibAnnote {#1} {#2} {\input{#2}}}{}%
}%
\providecommand \typeout [0]{\immediate \write \m@ne }%
\providecommand \selectlanguage [0]{\@gobble}%
\providecommand \bibinfo [0]{\@secondoftwo}%
\providecommand \bibfield [0]{\@secondoftwo}%
\providecommand \translation [1]{[#1]}%
\providecommand \BibitemOpen[0]{}%
\providecommand \bibitemStop [0]{}%
\providecommand \bibitemNoStop [0]{.\EOS\space}%
\providecommand \EOS [0]{\spacefactor3000\relax}%
\providecommand \BibitemShut [1]{\csname bibitem#1\endcsname}%
\bibitem{Hahn1950}%
  \BibitemOpen
  \bibfield{author}{%
  \bibinfo {author} {\bibfnamefont{E.~L.}\ \bibnamefont{Hahn}},\ }%
  \bibfield{journal}{%
  \Doi{10.1103/PhysRev.80.580}{\bibinfo {journal} {Phys. Rev.}}\ }%
  \textbf{\bibinfo {volume} {80}},\ \bibinfo {pages} {580} (\bibinfo {month}
  {Nov}\ \bibinfo {year} {1950}),\
  \url{http://link.aps.org/doi/10.1103/PhysRev.80.580}%
  \bibAnnoteFile{NoStop}{Hahn1950}%
\bibitem{Ott2008}%
  \BibitemOpen
  \bibfield{author}{%
  \bibinfo {author} {\bibfnamefont{E.}~\bibnamefont{Ott}}, \bibinfo {author}
  {\bibfnamefont{J.~H.}\ \bibnamefont{Platig}}, \bibinfo {author}
  {\bibfnamefont{T.~M.}\ \bibnamefont{Antonsen}},\ and\ \bibinfo {author}
  {\bibfnamefont{M.}~\bibnamefont{Girvan}},\ }%
  \bibfield{journal}{%
  \Doi{10.1063/1.2973816}{\bibinfo {journal} {Chaos: An Interdisciplinary
  Journal of Nonlinear Science}}\ }%
  \textbf{\bibinfo {volume} {18}},\ \bibinfo {eid} {037115} (\bibinfo {year}
  {2008}),\ \url{http://link.aip.org/link/?CHA/18/037115/1}%
  \bibAnnoteFile{NoStop}{Ott2008}%
\bibitem{Gould1967}%
  \BibitemOpen
  \bibfield{author}{%
  \bibinfo {author} {\bibfnamefont{R.~W.}\ \bibnamefont{Gould}}, \bibinfo
  {author} {\bibfnamefont{T.~M.}\ \bibnamefont{O'Neil}},\ and\ \bibinfo
  {author} {\bibfnamefont{J.~H.}\ \bibnamefont{Malmberg}},\ }%
  \bibfield{journal}{%
  \Doi{10.1103/PhysRevLett.19.219}{\bibinfo {journal} {Phys. Rev. Lett.}}\ }%
  \textbf{\bibinfo {volume} {19}},\ \bibinfo {pages} {219} (\bibinfo {month}
  {Jul}\ \bibinfo {year} {1967}),\
  \url{http://link.aps.org/doi/10.1103/PhysRevLett.19.219}%
  \bibAnnoteFile{NoStop}{Gould1967}%
\bibitem{Malmberg1968}%
  \BibitemOpen
  \bibfield{author}{%
  \bibinfo {author} {\bibfnamefont{J.~H.}\ \bibnamefont{Malmberg}}, \bibinfo
  {author} {\bibfnamefont{C.~B.}\ \bibnamefont{Wharton}}, \bibinfo {author}
  {\bibfnamefont{R.~W.}\ \bibnamefont{Gould}},\ and\ \bibinfo {author}
  {\bibfnamefont{T.~M.}\ \bibnamefont{O'Neil}},\ }%
  \bibfield{journal}{%
  \Doi{10.1103/PhysRevLett.20.95}{\bibinfo {journal} {Phys. Rev. Lett.}}\ }%
  \textbf{\bibinfo {volume} {20}},\ \bibinfo {pages} {95} (\bibinfo {month}
  {Jan}\ \bibinfo {year} {1968}),\
  \url{http://link.aps.org/doi/10.1103/PhysRevLett.20.95}%
  \bibAnnoteFile{NoStop}{Malmberg1968}%
\bibitem{Ott1970}%
  \BibitemOpen
  \bibfield{author}{%
  \bibinfo {author} {\bibfnamefont{E.}~\bibnamefont{Ott}},\ }%
  \bibfield{journal}{%
  \Doi{10.1017/S002237780000516X}{\bibinfo {journal} {Journal of Plasma
  Physics}}\ }%
  \textbf{\bibinfo {volume} {4}},\ \bibinfo {pages} {471} (\bibinfo {year}
  {1970}),\ \url{http://dx.doi.org/10.1017/S002237780000516X}%
  \bibAnnoteFile{NoStop}{Ott1970}%
\bibitem{Morigi2002}%
  \BibitemOpen
  \bibfield{author}{%
  \bibinfo {author} {\bibfnamefont{G.}~\bibnamefont{Morigi}}, \bibinfo {author}
  {\bibfnamefont{E.}~\bibnamefont{Solano}}, \bibinfo {author}
  {\bibfnamefont{B.-G.}\ \bibnamefont{Englert}},\ and\ \bibinfo {author}
  {\bibfnamefont{H.}~\bibnamefont{Walther}},\ }%
  \bibfield{journal}{%
  \Doi{10.1103/PhysRevA.65.040102}{\bibinfo {journal} {Phys. Rev. A}}\ }%
  \textbf{\bibinfo {volume} {65}},\ \bibinfo {pages} {040102} (\bibinfo {month}
  {Apr}\ \bibinfo {year} {2002}),\
  \url{http://link.aps.org/doi/10.1103/PhysRevA.65.040102}%
  \bibAnnoteFile{NoStop}{Morigi2002}%
\bibitem{Meunier2005}%
  \BibitemOpen
  \bibfield{author}{%
  \bibinfo {author} {\bibfnamefont{T.}~\bibnamefont{Meunier}}, \bibinfo
  {author} {\bibfnamefont{S.}~\bibnamefont{Gleyzes}}, \bibinfo {author}
  {\bibfnamefont{P.}~\bibnamefont{Maioli}}, \bibinfo {author}
  {\bibfnamefont{A.}~\bibnamefont{Auffeves}}, \bibinfo {author}
  {\bibfnamefont{G.}~\bibnamefont{Nogues}}, \bibinfo {author}
  {\bibfnamefont{M.}~\bibnamefont{Brune}}, \bibinfo {author}
  {\bibfnamefont{J.~M.}\ \bibnamefont{Raimond}},\ and\ \bibinfo {author}
  {\bibfnamefont{S.}~\bibnamefont{Haroche}},\ }%
  \bibfield{journal}{%
  \Doi{10.1103/PhysRevLett.94.010401}{\bibinfo {journal} {Phys. Rev. Lett.}}\
  }%
  \textbf{\bibinfo {volume} {94}},\ \bibinfo {pages} {010401} (\bibinfo {month}
  {Jan}\ \bibinfo {year} {2005}),\
  \url{http://link.aps.org/doi/10.1103/PhysRevLett.94.010401}%
  \bibAnnoteFile{NoStop}{Meunier2005}%
\bibitem{Bulatov1998}%
  \BibitemOpen
  \bibfield{author}{%
  \bibinfo {author} {\bibfnamefont{A.}~\bibnamefont{Bulatov}}, \bibinfo
  {author} {\bibfnamefont{A.}~\bibnamefont{Kuklov}}, \bibinfo {author}
  {\bibfnamefont{B.~E.}\ \bibnamefont{Vugmeister}},\ and\ \bibinfo {author}
  {\bibfnamefont{H.}~\bibnamefont{Rabitz}},\ }%
  \bibfield{journal}{%
  \Doi{10.1103/PhysRevA.57.3788}{\bibinfo {journal} {Phys. Rev. A}}\ }%
  \textbf{\bibinfo {volume} {57}},\ \bibinfo {pages} {3788} (\bibinfo {month}
  {May}\ \bibinfo {year} {1998}),\
  \url{http://link.aps.org/doi/10.1103/PhysRevA.57.3788}%
  \bibAnnoteFile{NoStop}{Bulatov1998}%
\bibitem{Piovella2003}%
  \BibitemOpen
  \bibfield{author}{%
  \bibinfo {author} {\bibfnamefont{N.}~\bibnamefont{Piovella}}, \bibinfo
  {author} {\bibfnamefont{V.}~\bibnamefont{Beretta}}, \bibinfo {author}
  {\bibfnamefont{G.~R.~M.}\ \bibnamefont{Robb}},\ and\ \bibinfo {author}
  {\bibfnamefont{R.}~\bibnamefont{Bonifacio}},\ }%
  \bibfield{journal}{%
  \Doi{10.1103/PhysRevA.68.021801}{\bibinfo {journal} {Phys. Rev. A}}\ }%
  \textbf{\bibinfo {volume} {68}},\ \bibinfo {pages} {021801} (\bibinfo {month}
  {Aug}\ \bibinfo {year} {2003}),\
  \url{http://link.aps.org/doi/10.1103/PhysRevA.68.021801}%
  \bibAnnoteFile{NoStop}{Piovella2003}%
\bibitem{Buchkremer2000}%
  \BibitemOpen
  \bibfield{author}{%
  \bibinfo {author} {\bibfnamefont{F.~B.~J.}\ \bibnamefont{Buchkremer}},
  \bibinfo {author} {\bibfnamefont{R.}~\bibnamefont{Dumke}}, \bibinfo {author}
  {\bibfnamefont{H.}~\bibnamefont{Levsen}}, \bibinfo {author}
  {\bibfnamefont{G.}~\bibnamefont{Birkl}},\ and\ \bibinfo {author}
  {\bibfnamefont{W.}~\bibnamefont{Ertmer}},\ }%
  \bibfield{journal}{%
  \Doi{10.1103/PhysRevLett.85.3121}{\bibinfo {journal} {Phys. Rev. Lett.}}\ }%
  \textbf{\bibinfo {volume} {85}},\ \bibinfo {pages} {3121} (\bibinfo {month}
  {Oct}\ \bibinfo {year} {2000}),\
  \url{http://link.aps.org/doi/10.1103/PhysRevLett.85.3121}%
  \bibAnnoteFile{NoStop}{Buchkremer2000}%
\bibitem{Maneshi2008}%
  \BibitemOpen
  \bibfield{author}{%
  \bibinfo {author} {\bibfnamefont{S.}~\bibnamefont{Maneshi}}, \bibinfo
  {author} {\bibfnamefont{J.~F.}\ \bibnamefont{Kanem}}, \bibinfo {author}
  {\bibfnamefont{C.}~\bibnamefont{Zhuang}}, \bibinfo {author}
  {\bibfnamefont{M.}~\bibnamefont{Partlow}},\ and\ \bibinfo {author}
  {\bibfnamefont{A.~M.}\ \bibnamefont{Steinberg}},\ }%
  \bibfield{journal}{%
  \Doi{10.1103/PhysRevA.77.022303}{\bibinfo {journal} {Phys. Rev. A}}\ }%
  \textbf{\bibinfo {volume} {77}},\ \bibinfo {pages} {022303} (\bibinfo {month}
  {Feb}\ \bibinfo {year} {2008}),\
  \url{http://link.aps.org/doi/10.1103/PhysRevA.77.022303}%
  \bibAnnoteFile{NoStop}{Maneshi2008}%
\bibitem{Andersen2003}%
  \BibitemOpen
  \bibfield{author}{%
  \bibinfo {author} {\bibfnamefont{M.~F.}\ \bibnamefont{Andersen}}, \bibinfo
  {author} {\bibfnamefont{A.}~\bibnamefont{Kaplan}},\ and\ \bibinfo {author}
  {\bibfnamefont{N.}~\bibnamefont{Davidson}},\ }%
  \bibfield{journal}{%
  \Doi{10.1103/PhysRevLett.90.023001}{\bibinfo {journal} {Phys. Rev. Lett.}}\
  }%
  \textbf{\bibinfo {volume} {90}},\ \bibinfo {pages} {023001} (\bibinfo {month}
  {Jan}\ \bibinfo {year} {2003}),\
  \url{http://link.aps.org/doi/10.1103/PhysRevLett.90.023001}%
  \bibAnnoteFile{NoStop}{Andersen2003}%
\bibitem{Elyutin2005}%
  \BibitemOpen
  \bibfield{author}{%
  \bibinfo {author} {\bibfnamefont{S.}~\bibnamefont{Elyutin}},\ }%
  \bibfield{journal}{%
  \bibinfo {journal} {Optics and Spectroscopy}\ }%
  \textbf{\bibinfo {volume} {98}},\ \bibinfo {pages} {605} (\bibinfo {year}
  {2005}),\ ISSN \bibinfo {issn} {0030-400X},\ \bibinfo {note}
  {10.1134/1.1914901},\ \url{http://dx.doi.org/10.1134/1.1914901}%
  \bibAnnoteFile{NoStop}{Elyutin2005}%
\bibitem{Raithel1998}%
  \BibitemOpen
  \bibfield{author}{%
  \bibinfo {author} {\bibfnamefont{G.}~\bibnamefont{Raithel}}, \bibinfo
  {author} {\bibfnamefont{W.~D.}\ \bibnamefont{Phillips}},\ and\ \bibinfo
  {author} {\bibfnamefont{S.~L.}\ \bibnamefont{Rolston}},\ }%
  \bibfield{journal}{%
  \Doi{10.1103/PhysRevLett.81.3615}{\bibinfo {journal} {Phys. Rev. Lett.}}\ }%
  \textbf{\bibinfo {volume} {81}},\ \bibinfo {pages} {3615} (\bibinfo {month}
  {Oct}\ \bibinfo {year} {1998}),\
  \url{http://link.aps.org/doi/10.1103/PhysRevLett.81.3615}%
  \bibAnnoteFile{NoStop}{Raithel1998}%
\bibitem{Pitaevskii1997}%
  \BibitemOpen
  \bibfield{author}{%
  \bibinfo {author} {\bibfnamefont{L.}~\bibnamefont{Pitaevskii}},\ }%
  \bibfield{journal}{%
  \Doi{10.1016/S0375-9601(97)00261-2}{\bibinfo {journal} {Physics Letters A}}\
  }%
  \textbf{\bibinfo {volume} {229}},\ \bibinfo {pages} {406} (\bibinfo {year}
  {1997}),\ ISSN \bibinfo {issn} {0375-9601},\
  \url{http://www.sciencedirect.com/science/article/pii/S0375960197002612}%
  \bibAnnoteFile{NoStop}{Pitaevskii1997}%
\bibitem{Robinett2004}%
  \BibitemOpen
  \bibfield{author}{%
  \bibinfo {author} {\bibfnamefont{R.}~\bibnamefont{Robinett}},\ }%
  \bibfield{journal}{%
  \Doi{10.1016/j.physrep.2003.11.002}{\bibinfo {journal} {Physics Reports}}\ }%
  \textbf{\bibinfo {volume} {392}},\ \bibinfo {pages} {1} (\bibinfo {year}
  {2004}),\ ISSN \bibinfo {issn} {0370-1573},\
  \url{http://www.sciencedirect.com/science/article/pii/S0370157303004381}%
  \bibAnnoteFile{NoStop}{Robinett2004}%
\bibitem{Eberly1980}%
  \BibitemOpen
  \bibfield{author}{%
  \bibinfo {author} {\bibfnamefont{J.~H.}\ \bibnamefont{Eberly}}, \bibinfo
  {author} {\bibfnamefont{N.~B.}\ \bibnamefont{Narozhny}},\ and\ \bibinfo
  {author} {\bibfnamefont{J.~J.}\ \bibnamefont{Sanchez-Mondragon}},\ }%
  \bibfield{journal}{%
  \Doi{10.1103/PhysRevLett.44.1323}{\bibinfo {journal} {Phys. Rev. Lett.}}\ }%
  \textbf{\bibinfo {volume} {44}},\ \bibinfo {pages} {1323} (\bibinfo {month}
  {May}\ \bibinfo {year} {1980}),\
  \url{http://link.aps.org/doi/10.1103/PhysRevLett.44.1323}%
  \bibAnnoteFile{NoStop}{Eberly1980}%
\bibitem{Buzek1995}%
  \BibitemOpen
  \bibfield{author}{%
  \bibinfo {author} {\bibfnamefont{V.}~\bibnamefont{Bu\v{z}ek}}\ and\ \bibinfo
  {author} {\bibfnamefont{P.}~\bibnamefont{Knight}},\ }%
  \enquote{\bibinfo {title} {I: Quantum interference, superposition states of
  light, and nonclassical effects},}\ in\ \emph{\bibinfo {booktitle} {Progress
  in Optics}},\ Vol.~\bibinfo {volume} {34},\ \bibinfo {editor} {edited by\
  \bibinfo {editor} {\bibfnamefont{E.}~\bibnamefont{Wolf}}}\ (\bibinfo
  {publisher} {Elsevier},\ \bibinfo {year} {1995})\ pp.\ \bibinfo {pages}
  {1--158},\
  \url{http://www.sciencedirect.com/science/article/pii/S007966380870324X}%
  \bibAnnoteFile{NoStop}{Buzek1995}%
\bibitem{Meekhof1996}%
  \BibitemOpen
  \bibfield{author}{%
  \bibinfo {author} {\bibfnamefont{D.~M.}\ \bibnamefont{Meekhof}}, \bibinfo
  {author} {\bibfnamefont{C.}~\bibnamefont{Monroe}}, \bibinfo {author}
  {\bibfnamefont{B.~E.}\ \bibnamefont{King}}, \bibinfo {author}
  {\bibfnamefont{W.~M.}\ \bibnamefont{Itano}},\ and\ \bibinfo {author}
  {\bibfnamefont{D.~J.}\ \bibnamefont{Wineland}},\ }%
  \bibfield{journal}{%
  \Doi{10.1103/PhysRevLett.76.1796}{\bibinfo {journal} {Phys. Rev. Lett.}}\ }%
  \textbf{\bibinfo {volume} {76}},\ \bibinfo {pages} {1796} (\bibinfo {month}
  {Mar}\ \bibinfo {year} {1996}),\
  \url{http://link.aps.org/doi/10.1103/PhysRevLett.76.1796}%
  \bibAnnoteFile{NoStop}{Meekhof1996}%
\bibitem{Brune1996}%
  \BibitemOpen
  \bibfield{author}{%
  \bibinfo {author} {\bibfnamefont{M.}~\bibnamefont{Brune}}, \bibinfo {author}
  {\bibfnamefont{F.}~\bibnamefont{Schmidt-Kaler}}, \bibinfo {author}
  {\bibfnamefont{A.}~\bibnamefont{Maali}}, \bibinfo {author}
  {\bibfnamefont{J.}~\bibnamefont{Dreyer}}, \bibinfo {author}
  {\bibfnamefont{E.}~\bibnamefont{Hagley}}, \bibinfo {author}
  {\bibfnamefont{J.~M.}\ \bibnamefont{Raimond}},\ and\ \bibinfo {author}
  {\bibfnamefont{S.}~\bibnamefont{Haroche}},\ }%
  \bibfield{journal}{%
  \Doi{10.1103/PhysRevLett.76.1800}{\bibinfo {journal} {Phys. Rev. Lett.}}\ }%
  \textbf{\bibinfo {volume} {76}},\ \bibinfo {pages} {1800} (\bibinfo {month}
  {Mar}\ \bibinfo {year} {1996}),\
  \url{http://link.aps.org/doi/10.1103/PhysRevLett.76.1800}%
  \bibAnnoteFile{NoStop}{Brune1996}%
\bibitem{Hofheinz2008}%
  \BibitemOpen
  \bibfield{author}{%
  \bibinfo {author} {\bibfnamefont{M.}~\bibnamefont{{Hofheinz}}}, \bibinfo
  {author} {\bibfnamefont{E.~M.}\ \bibnamefont{{Weig}}}, \bibinfo {author}
  {\bibfnamefont{M.}~\bibnamefont{{Ansmann}}}, \bibinfo {author}
  {\bibfnamefont{R.~C.}\ \bibnamefont{{Bialczak}}}, \bibinfo {author}
  {\bibfnamefont{E.}~\bibnamefont{{Lucero}}}, \bibinfo {author}
  {\bibfnamefont{M.}~\bibnamefont{{Neeley}}}, \bibinfo {author}
  {\bibfnamefont{A.~D.}\ \bibnamefont{{O'Connell}}}, \bibinfo {author}
  {\bibfnamefont{H.}~\bibnamefont{{Wang}}}, \bibinfo {author}
  {\bibfnamefont{J.~M.}\ \bibnamefont{{Martinis}}},\ and\ \bibinfo {author}
  {\bibfnamefont{A.~N.}\ \bibnamefont{{Cleland}}},\ }%
  \bibfield{journal}{%
  \Doi{10.1038/nature07136}{\bibinfo {journal} {\nat}}\ }%
  \textbf{\bibinfo {volume} {454}},\ \bibinfo {pages} {310} (\bibinfo {month}
  {Jul.}\ \bibinfo {year} {2008}),\
  \url{http://dx.doi.org/10.1038/nature07136}%
  \bibAnnoteFile{NoStop}{Hofheinz2008}%
\bibitem{Sakurai1994}%
  \BibitemOpen
  \bibfield{author}{%
  \bibinfo {author} {\bibfnamefont{J.~J.}\ \bibnamefont{Sakurai}},\ }%
  \emph{\bibinfo {title} {Modern Quantum Mechanics}},\ \bibinfo {edition}
  {{R}evised}\ ed.\ (\bibinfo {publisher} {Addison-Wesley Publishing Company},\
  \bibinfo {address} {Reading, MA},\ \bibinfo {year} {1994})%
  \bibAnnoteFile{NoStop}{Sakurai1994}%
\bibitem{Gardiner2000}%
  \BibitemOpen
  \bibfield{author}{%
  \bibinfo {author} {\bibfnamefont{S.~A.}\ \bibnamefont{Gardiner}}, \bibinfo
  {author} {\bibfnamefont{D.}~\bibnamefont{Jaksch}}, \bibinfo {author}
  {\bibfnamefont{R.}~\bibnamefont{Dum}}, \bibinfo {author}
  {\bibfnamefont{J.~I.}\ \bibnamefont{Cirac}},\ and\ \bibinfo {author}
  {\bibfnamefont{P.}~\bibnamefont{Zoller}},\ }%
  \bibfield{journal}{%
  \Doi{10.1103/PhysRevA.62.023612}{\bibinfo {journal} {Phys. Rev. A}}\ }%
  \textbf{\bibinfo {volume} {62}},\ \bibinfo {pages} {023612} (\bibinfo {month}
  {Jul}\ \bibinfo {year} {2000}),\
  \url{http://link.aps.org/doi/10.1103/PhysRevA.62.023612}%
  \bibAnnoteFile{NoStop}{Gardiner2000}%
\bibitem{Kasperkovitz1995}%
  \BibitemOpen
  \bibfield{author}{%
  \bibinfo {author} {\bibfnamefont{P.}~\bibnamefont{Kasperkovitz}}\ and\
  \bibinfo {author} {\bibfnamefont{M.}~\bibnamefont{Peev}},\ }%
  \bibfield{journal}{%
  \Doi{10.1103/PhysRevLett.75.990}{\bibinfo {journal} {Phys. Rev. Lett.}}\ }%
  \textbf{\bibinfo {volume} {75}},\ \bibinfo {pages} {990} (\bibinfo {month}
  {Aug}\ \bibinfo {year} {1995}),\
  \url{http://link.aps.org/doi/10.1103/PhysRevLett.75.990}%
  \bibAnnoteFile{NoStop}{Kasperkovitz1995}%
\bibitem{Manfredi1996}%
  \BibitemOpen
  \bibfield{author}{%
  \bibinfo {author} {\bibfnamefont{G.}~\bibnamefont{Manfredi}}\ and\ \bibinfo
  {author} {\bibfnamefont{M.~R.}\ \bibnamefont{Feix}},\ }%
  \bibfield{journal}{%
  \Doi{10.1103/PhysRevE.53.6460}{\bibinfo {journal} {Phys. Rev. E}}\ }%
  \textbf{\bibinfo {volume} {53}},\ \bibinfo {pages} {6460} (\bibinfo {month}
  {Jun}\ \bibinfo {year} {1996}),\
  \url{http://link.aps.org/doi/10.1103/PhysRevE.53.6460}%
  \bibAnnoteFile{NoStop}{Manfredi1996}%
\bibitem{Gross1961}%
  \BibitemOpen
  \bibfield{author}{%
  \bibinfo {author} {\bibfnamefont{E.}~\bibnamefont{Gross}},\ }%
  \bibfield{journal}{%
  \bibinfo {journal} {Il Nuovo Cimento (1955-1965)}\ }%
  \textbf{\bibinfo {volume} {20}},\ \bibinfo {pages} {454} (\bibinfo {year}
  {1961}),\ ISSN \bibinfo {issn} {1827-6121},\ \bibinfo {note}
  {10.1007/BF02731494},\ \url{http://dx.doi.org/10.1007/BF02731494}%
  \bibAnnoteFile{NoStop}{Gross1961}%
\bibitem{Pitaevskii1961}%
  \BibitemOpen
  \bibfield{author}{%
  \bibinfo {author} {\bibfnamefont{L.~P.}\ \bibnamefont{Pitaevskii}},\ }%
  \bibfield{journal}{%
  \bibinfo {journal} {Soviet Phys. {\rm JETP}}\ }%
  \textbf{\bibinfo {volume} {13}},\ \bibinfo {pages} {451} (\bibinfo {year}
  {1961})%
  \bibAnnoteFile{NoStop}{Pitaevskii1961}%
\bibitem{Pethick2008}%
  \BibitemOpen
  \bibfield{author}{%
  \bibinfo {author} {\bibfnamefont{C.~J.}\ \bibnamefont{Pethick}}\ and\
  \bibinfo {author} {\bibfnamefont{H.}~\bibnamefont{Smith}},\ }%
  \emph{\bibinfo {title} {Bose-Einstein Condensation in Dilute Gases; 2nd
  ed.}}\ (\bibinfo {publisher} {Cambridge Univ. Press},\ \bibinfo {address}
  {Cambridge},\ \bibinfo {year} {2008})%
  \bibAnnoteFile{NoStop}{Pethick2008}%
\bibitem{Moulieras2012}%
  \BibitemOpen
  \bibfield{author}{%
  \bibinfo {author} {\bibfnamefont{S.}~\bibnamefont{Moulieras}}, \bibinfo
  {author} {\bibfnamefont{A.~G.}\ \bibnamefont{Monastra}}, \bibinfo {author}
  {\bibfnamefont{M.}~\bibnamefont{Saraceno}},\ and\ \bibinfo {author}
  {\bibfnamefont{P.}~\bibnamefont{Leboeuf}},\ }%
  \bibfield{journal}{%
  \Doi{10.1103/PhysRevA.85.013841}{\bibinfo {journal} {Phys. Rev. A}}\ }%
  \textbf{\bibinfo {volume} {85}},\ \bibinfo {pages} {013841} (\bibinfo {month}
  {Jan}\ \bibinfo {year} {2012}),\
  \url{http://link.aps.org/doi/10.1103/PhysRevA.85.013841}%
  \bibAnnoteFile{NoStop}{Moulieras2012}%
\end{thebibliography}
\end{document}